\documentclass[fleqn,10pt]{wlscirep}
\usepackage{amsmath,amsfonts,amssymb}
\usepackage{graphicx}
\usepackage{dcolumn}
\usepackage{bm}
\usepackage{dcolumn}

\title{Change in the magnetic configurations of tubular nanostructures by tuning dipolar interactions}
\author{H. D. Salinas}
\affil{Grupo de Magnetismo y Simulaci\'on G+, Instituto de
  F\'isica, Universidad de Antioquia. A.A. 1226, Medell\'in, Colombia, hernan.salinas@udea.edu.co}

\author{J. Restrepo}
\affil{Grupo de Magnetismo y Simulaci\'on G+, Instituto de
  F\'isica, Universidad de Antioquia. A.A. 1226, Medell\'in, Colombia, johans.restrepo@udea.edu.co}

\author{\`Oscar Iglesias} 
\affil{Departament de F\'isica de la Matèria Condensada and Institut de
  Nanoci\`encia i Nanotecnologia, Universitat de Barcelona, Av. Diagonal
  647, 08028 Barcelona, Spain, oscar@ffn.ub.es}

\date{\today}

\begin{abstract}
We have investigated the equilibrium states of ferromagnetic single wall nanotubes  by means of atomistic Monte Carlo simulations of a zig-zag lattice of Heisenberg spins on the surface of a cylinder.
The main focus of our study is to determine how the competition between short-range exchange ($J$) and long-range dipolar ($D$) interactions influences the low temperature magnetic order of the nanotubes as well as the thermal-driven transitions involved.
Apart from the uniform and vortex states occurring for dominant $J$ or $D$, we find that helical states become stable for a range of intermediate values of $\gamma=D/J$ that depends on the radius and length of the nanotube.
Introducing a vorticity order parameter to better characterize helical and vortex states, we find the pseudo-critical temperatures for the transitions between these states and we establish the magnetic phase diagrams of their stability regions as a function of the nanotube aspect ratio. 
Comparison of the energy of the states obtained by simulation with those of simpler theoretical structures that interpolate continuously between them, reveals a high degree of metastability of the helical structures that might be relevant for their reversal modes.    
\end{abstract}

\begin{document}
\flushbottom
\maketitle
%
%
\thispagestyle{empty}

\section*{Introduction}

When planar structures are extended to 3D by changing their curvature, the magnetic behavior of the corresponding surface can change significantly, leading for example to the appearance of new magnetic configurations that are not energetically favorable in the planar case \cite{Fernandez-Pacheco_natcomm2016,Hertel_jpcm2016} and also to the induction of anisotropic and chiral effects that merely originate from the curvature of the surface \cite{Sander_jpd2017,Sloika_jmmm2017,Streubel_jpd2016,Hertel_spin2013,Sheka_prb2015}. 
A nice example of these, are tubular magnetic nanostructures that have gained recent interest because the increased functionality arising from their inner and outer surfaces may contribute to improve existing applications \cite{Fan_acsnano2009,Fernandez-Pacheco_natcomm2016} based on other proposals. 
This fact makes them very attractive for diverse applications in several fields involving biomedicine \cite{Xie_nanotec2008,Pantarotto_chembio2013}, spintronics \cite{Krusin-Elbaum_nature2004}, magnetic sensors or high-density magnetic storage devices \cite{Parkin_science2008} among others. 
The hollow nature of nanotubes favors the formation of closed flux, core free configurations. These kind of states becomes particularly relevant for magnetic storage purposes because they do not produce stray fields (flux-closure configuration) avoiding the consequent leakage of magnetic flux spreading outward from the tube. Additionally, the core-free aspect offers advantages over nanowires, as it allows more controllable reversal processes, increased stability \cite{Yan_apl2011}, and faster domain wall propagation \cite{Wang_prl2005,Landeros_prb2009}.

The rapid advance in fabrication techniques available nowadays allows to synthesize tubular nanostrutures by following different routes like chemical or electro-deposition \cite{Nielsch_jap2005,Kohli_jppc2010},  hydrogen reduction \cite{Siu_apl2004}, sol-gel \cite{Xu_jpd2008}, electrochemical synthesis \cite{Proenca_book2015}, Kirkendall effect \cite{Wang_cc2010}, self-assembly of NPs \cite{Harada_pccp2010}and others \cite{Ye_crssms2012} with a high degree of control on their geometrical parameters. 
In particular, different realizations of magnetic nanotubular structures have been achieved in the form of rolled up magnetic nanomembranes \cite{Streubel_NanoLetters2014}, magnetite nanotubes on MgO wires \cite{Liu_jacs2005}, nanoparticles deposited on the surface of nanotubes \cite{Sotnikov_prb2017}, or carbon nanotubes either filled with magnetic nanoparticles (NP) \cite{Jang_advmat2003,Gao_jpcb2006, Lamanna_nanoscale2013} or with NP and networks of molecular molecules or nanoparticles attached to their surface \cite{Kim_jpcc2010, Bogani_Angewandte2009}, only to mention some examples.

Vortex states having the magnetic moments circulating around the symmetry axis are expected to be formed in structures having cylindrical symmetry \cite{Cowburn_prl1999}. As predicted by theoretical studies, vortex and helical states have been found giving rise to a rich magnetic phase diagram where both temperature and geometry (aspect ratio) play an important role \cite{Chen_jmmm2007,Escrig_jmmm2007}. However, since their magnetization is close to zero, standard magnetic characterization precluded their experimental evidence until the advent of more sensitive techniques allowed their observation in individual nanocylinders, nanotubes and dot-shaped elements of different compositions by means of electron holography \cite{Raabe_jap2000,Biziere_Nanolett2013}, cantilever-based torque magnetometry \cite{Weber_Nanolett2012,Wyss_prb2017,Vasyukov_NanoLett2018,Mehlin_prb2018}, nanoscale torque-SQUID magnetometry \cite{Buchter_prl2013}, spin-polarized scanning tunneling spectroscopy \cite{Yamasaki_prl2003}, X-ray microscopy \cite{Zimmermann_nanolett2018}  and MFM \cite{Lebib_jap2001}. 

Different theoretical and simulation approaches have been used in order to study the thermodynamic properties of magnetic hollow nanocylinders, including micromagnetic studies \cite{Landeros_prb2009,Allende_ejpb2008}, scaling techniques \cite{Albuquerque_prl2002,Landeros_prb2005}, many-body Green's functions  \cite{Mi_jmmm2010,Mi_jmmm2016,Mi_phyb2016}, ab-initio calculations \cite{Lehtinen_prl2004,Shimada_prb2011,Shimada_nanolett2013}, and Monte Carlo (MC) simulations \cite{Konstantinova_jmmm2008}.

Most of these works, have focused on the study of magnetic configurations as a function of the geometric parameters of the nanotubes for given values of material parameters. However, the occurrence of such phases and their thermal stability becomes also determined by the competition between the short-range exchange  and long-range magnetic dipolar interactions\cite{Chien_phystoday2007, Salinas_jsnm2012}, quantified by $\gamma$.
The value of $\gamma$ can be tuned not only by changing parameters of the material ($J$ and $\mu$) , but also by changing the lattice spacing between the magnetic entities (magnetic ions or nanoparticles) and their spatial arrangement, since both contribute to the dipolar interaction energy.  
Real systems in which this can be achieved are, for example, carbon nanotubes used as structural templates or guides to attach or decorate their surface with high-moment magnetic nanoparticles, clusters or single-molecule magnets  \cite{Masotti_ijms2013}.

Therefore, the present work will focus on studying the changes in the equilibrium configurations with $\gamma$ considering first a fixed tube geometry and then extending the calculations to a range of tube radii and lengths. 
To address the problem, we use standard Metropolis MC to study the finite-temperature properties, and we revert to analytical or numerical calculations when comparing the properties of ground-state configurations obtained from simulations to theoretical ones. 
It turns out that the different equilibrium states with well-defined chiral order have magnetization close to zero and, therefore, this magnitude is no longer a valid order parameter to characterize the transitions between the mentioned states. We will show that these states and the transitions in between can be correctly identified and characterized by introducing the rotational of the magnetization as a measure of vorticity or circularity of the moments around the tube. This, combined with a detailed inspection of the equilibrium spin patterns, will allow us to determine the magnetic phase diagrams of their stability regions as a function of the nanotube aspect ratio.

The outline of the manuscript is a follows. We start by presenting the model for the nanotubes and computational details of the simulations. Then, the results of the equilibrium properties are presented for a $(8,15)$ tube, followed by a detailed analysis of the ground state configurations. In the next subsection, we extend the calculations to tubes with different radii and lengths and present the corresponding phase diagrams as a function of the parameter $\gamma$. Finally, we analyze the metastability of the helical states by comparing the simulation results to analytical calculations based on two kind of states that can be theoretically parametrized. We finish with a discussion and final conclusions.

\section*{Results}
\subsection*{Model and Computational details}
\label{Model_Sec}

Our model system to simulate single-wall nanotubes consists of Heisenberg classical unitary spins  $ \vec{S}=(S_x,S_y,S_z)$  placed on the surface of a cylinder of radius $R$ and finite length $L$ \cite{Salinas_jsnm2012}. The structure is formed by rolling a planar squared spin lattice with lattice parameter $a$  along the (1,1) direction onto a cylindrical geometry in order to get a zig-zag ended tube that can be characterized by $N_z$ layers of rings stacked along the z axis with $N$ spins per layer. 
For a given $N$,  the tube is characterized by the angle between consecutive spins in the ring $\varphi_N= 2\pi/N$ , interlayer separation $l=\sqrt{2}a/2$ and radius $R=\frac{N \sqrt 2}{2\pi}a$.
Therefore,  the distance  between two consecutive spins in the ring is given by $d_1=R\sqrt{2(1-\cos\varphi_N)}=2R\sin (\varphi_N/2)$ (see Supplementary Fig. S1), and that between nn in adjacent rings is $a$. In what follows, we will designate the tubes by the notation $(N,N_z)$.

\begin{figure}[htb]
  \centering
\includegraphics[width=\columnwidth]{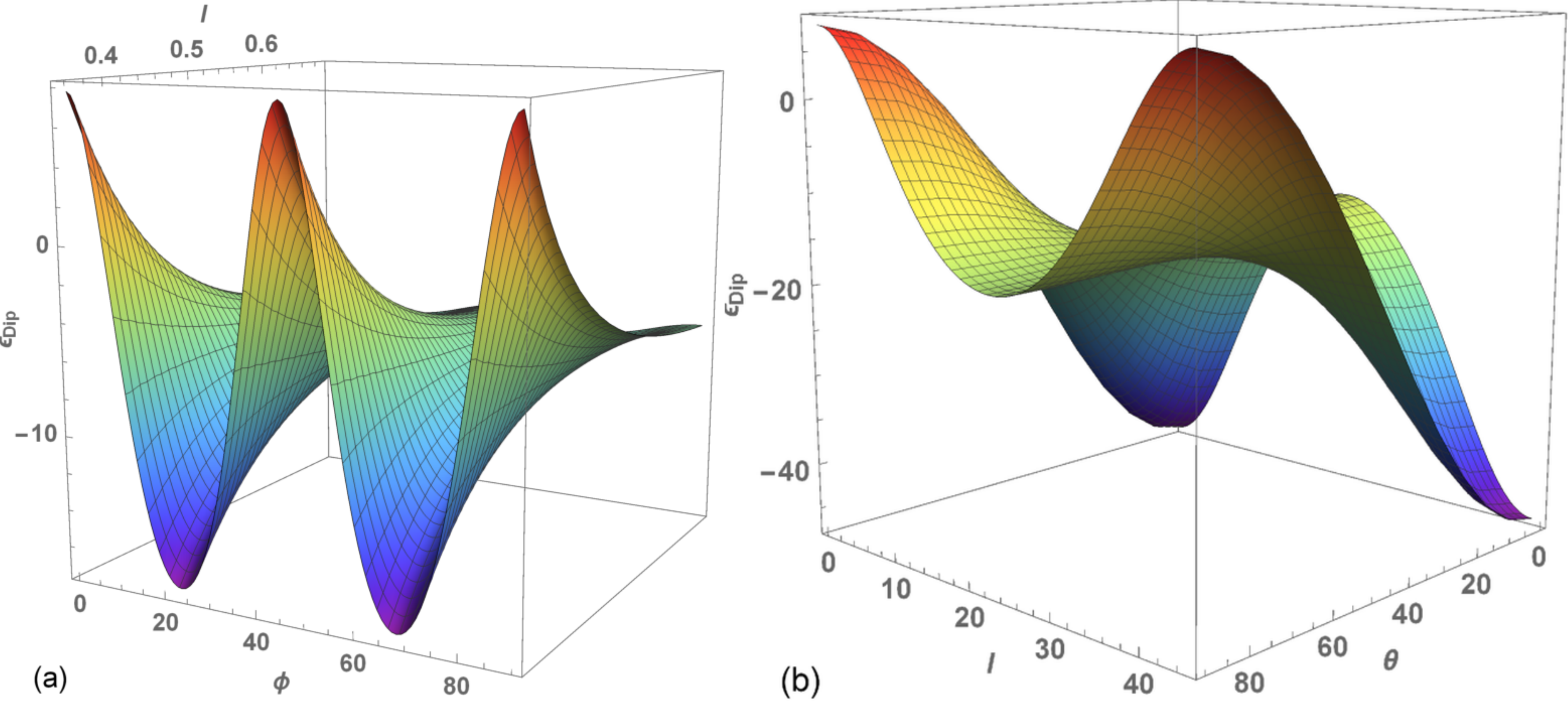}
\caption{Dipolar energy landscape for a (8,15) tube ($\varphi_1=\pi/4$) with spins aligned along the surface tangent  to the tube: (a) as a function of $\phi$, the angle between nn in consecutive rings, and interlayer separation $l$ , (b) as a function of $\theta$ (their common angle  with respect to the tube axis) and $\phi$, when the interlayer separation is fixed at $l=\sqrt{2}a/2$.}
\label{Fig_Energyscape}
\end{figure}

Of course, there are other ways to arrange spins on a regular lattice on the surface of a cylinder (for example, varying the interlayer separation and the kind of stacking in between layers), but not all of them will be stable from the point of view of dipolar energy alone \cite{Shimada_nanolett2013}. 
In order to justify the election of zig-zag tubes for our study, we have computed the dipolar energy of tubes with fixed $(N, N_z)$ and a spin configuration forming a perfect vortex. 
In Fig. \ref{Fig_Energyscape} (a), we show the energy landscape as a function of $l$ and $\phi$, the relative angle between consecutive layers. The dipolar energy presents minima at $\phi= \varphi_N/2$ and maxima at $\phi=0, \pi/4, \pi/2$ independently of $l$. This observation indicates that zig-zag tubes with spins in a vortex configuration are minimum energy states, whereas the formation of tubes with AA stacking (consecutive rings staked onto each other, with spins forming columns) is not energetically favored by dipolar interactions.  
Next, we fix $l$ to the minimum possible separation between rings and we allow the polar angle $\theta$ of the spin orientation to vary from parallel [ferromagnetic (FM) configuration, $\theta=0$] to perpendicular [vortex (V) configuration, $\theta=\pi/2$] to the tube axis. The computed energy landscape shown in Fig. \ref{Fig_Energyscape}(b) demonstrates that, depending on the degree of vertical misalignment between spins in consecutive rings, different spatial arrangements of the spins are favored by the dipolar interactions. Whereas for AA stacked tubes, alignment along the tube axis is preferred, for zig-zag tubes, a vortex configuration ($\theta=\pi/2$) has the minimum possible dipolar energy among all considered kinds of ordering.  

The spin interaction Hamiltonian can be written as
\begin{equation}
  {\cal H}=E_{ex}+E_{dip} \  ,
\end{equation}
where $E_{ex}$ stands for an isotropic short-range exchange coupling between nearest neighbors (nn) spins:
\begin{equation}
  E_{ex}=-\sum_{\langle i,j \rangle } J_{ij} \vec{S_i}\cdot\vec{S_j} \ ,
\end{equation}
being $J_ {ij} = J$ the positive exchange constant accounting for ferromagnetic interactions. $ E_{dip}$ stands for  the long-range magnetic dipolar interaction given by
\begin{equation}
  \label{Eq_Edip}
  E_{dip} = D\sum_{i< j}\left(\frac{
      \vec{S_{i}}\cdot\vec{S_{j}}-3(\vec{S_{i}}\cdot\hat{r}_{ij})(\vec{S_{j}}\cdot\hat{r}_{ij})} {|\vec{r}_{ij}|^3}\right),
\end{equation}
where summation expands over all the set of spin pairs, $D=\frac{\mu_{0}}{4\pi}\mu ^2/a^3$ is the dipolar coupling constant, with $\mu_{0} $ and $\mu$ the magnetic permeability of vacuum and the magnetic moment per spin respectively, and $\vec{r}_{ij}$ is the relative position vector between $i$ and $j$ spins.
We define the parameter $\gamma$ = $D/J$ in order to quantify the degree of competition between the short-range and long-range interactions. Temperature ($T$) is expressed in reduced $J/k_B$ units where $k_B$ is the Boltzmann constant. The Hamiltonian does not contain a magnetocrystalline anisotropic energy term as so far we are interested in soft magnetic materials where the associated energy can be neglected compared to the exchange and magnetostatic counterparts \cite{Li-Ying}. The total energy per spin can then be written as
\begin{equation}
  \label{Eq_TotalE}
  \epsilon_{total}/J=\epsilon_{ex}+\gamma \epsilon_{dip} \ .
\end{equation}

We have conducted standard Metropolis MC simulations to analyze the temperature dependence of thermodynamic observables and low temperature spin configurations arising from the competition between dipolar and exchange interactions. In order to be able to reach low temperature configurations close to the vicinity of the thermodynamic equilibrium for a given $\gamma$, we have employed the following protocols for the thermal dependence and spin updates:
(1) linear decrease of temperature with random homogeneous spin trials; (2) linear decrease of temperature with spin trials restricted to lie inside a cone with aperture varying with temperature; and (3) simulated annealing with a temperature power law decay of the form $T=T_0\alpha^k$ where spin updates are also restricted to lie inside a cone.
For the first two, initial and final temperatures are $50$ J/k$_B$, $0.2$ J/$k_B$ respectively, with a temperature step of $0.2$ J/k$_B$.
For protocol (3) simulations are started  at $T_0= 50$ J/k$_B$ and ended at $0.113$ J/k$_B$ after $200$ temperatures with $\alpha=0.97$. The number of MC steps used at each temperature is $5\times 10^{3}$, with $2\times 10^{3}$ used for averaging. Error bars in subsequent results have been obtained after averaging over $5$ independent simulations starting with different seeds.

\begin{figure}[htb]
  \centering
\includegraphics[width=0.8\columnwidth]{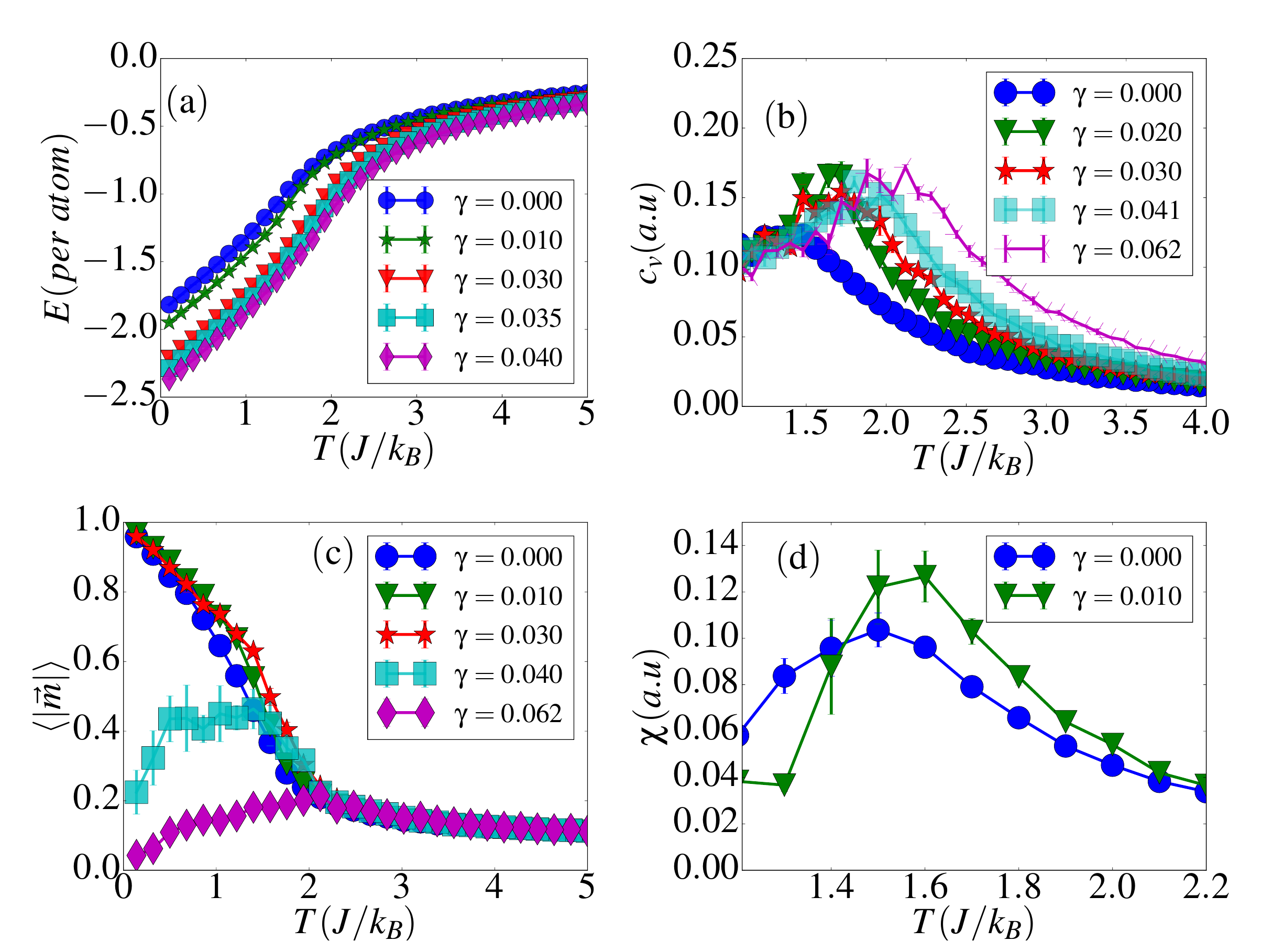}
\caption{Thermal dependence of energy (a), specific heat (b,c), magnetization (d) and susceptibility (e) for different values of $\gamma$ for a (8,15) tube. Error bars correspond to averages over $5$ independent runs.}
\label{Fig_Thermaldependence}
\end{figure}

\begin{figure}[htb]
  \centering
\includegraphics[width=0.8\columnwidth]{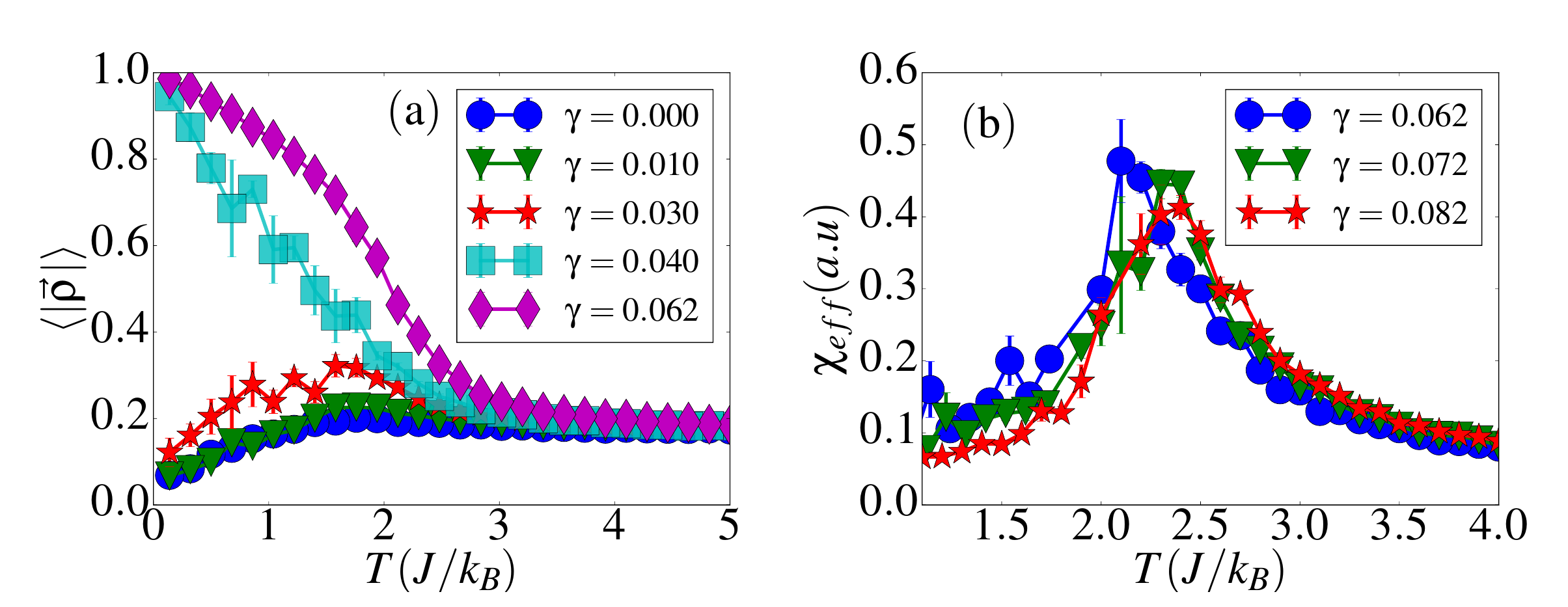}
\caption{Thermal dependence of the vorticity (a) and the associated susceptibility (b) for different values of $\gamma$ for a $(8,15)$ tube. $\langle \rho \rangle $ has been normalized to the maximum attainable value. Error bars correspond to averages over $5$ independent runs.}
\label{Fig_Thermaldependence2}
\end{figure}
In order to generate spin trials inside a cone, we generated a  random vector $\vec{v}_i$ inside a sphere of radius $S_{max}$ which is added to the current spin vector $\vec{S}_i$ and then the resulting vector $\vec{S}'_i$ is normalized, thus $\vec{S}'_i=\vec{v}_i+\vec{S}_i$. In order to keep the acceptance rates high, we use an adaptive scheme for the spin updates by reducing the aperture of the cone progressively as $T$ decreases. We initially set $S_{max}= 2$ at high $T$ so that initially the trial spin can reach any state within the Bloch sphere and then, at every temperature, we replace it by the maximum value accepted at trials during the previous temperature.  

\subsection*{Thermal dependence of near-equilibrium properties}
\label{Equil_Sec}

Simulation results for the thermal dependence of energy, specific heat, magnetic susceptibility and magnetization per site of a (8,15) nanotube are shown in Fig. \ref{Fig_Thermaldependence} for a range of $\gamma$ values between $0$ and $0.082$.  Energy curves shown in panel (a) display a concavity change pointing to a well-defined thermal-driven phase transition for all the values of $\gamma$ considered. For a given temperature, the energy decreases as $\gamma$ is increased independently of the temperature.

In particular, for $\gamma \leq 0.03$, the magnetization curves in Fig. \ref{Fig_Thermaldependence}(c) are smooth and well-behaved accounting for a well-defined ferromagnetic (FM) to paramagnetic (PM) transition above some pseudo-critical temperature $T_c$, below which the spins tend to be aligned along the tube axis. This is also signaled by the peak in the specific heat curves of Fig. \ref{Fig_Thermaldependence}(b) that appear rounded due to finite-size effects. 
Similar features can be found in the thermal dependence of the susceptibility as observed in Fig. \ref{Fig_Thermaldependence}(d), where we observe that the pseudo-critical temperature for the PM to FM transition (signaled by the position of the peak in the susceptibility curve) is shifted to higher values when dipolar interactions are turned on. 

However, for $\gamma\ge 0.04$,  the magnetization tends to zero when decreasing $T$ and it is no longer a valid order parameter since, as we will show latter,  the dominant dipolar interactions favor the appearance of closed flux states in the form of a vortex circulating around the tube axis.
In order to better characterize the onset of these vortex states, we have found convenient to define a new order parameter that we will call vorticity that characterizes the degree of circularity of a given magnetic structure, which is defined locally as the vector:
\begin{equation} 
  \vec{\rho_i}=\vec{\nabla}\times \vec{S}_i\ .
\end{equation}
Correspondingly, the average $z$ component of the vorticity of the whole tube can be written as a discrete sum over lattice points:
\begin{equation}
  \label{ec:vorticidad-rotacional}
  \langle  \rho^{z}\rangle =\frac{1}{NN_z}\sum_{k,l} \left(\frac{ (S_{k+1,l}^{y}- S_{k,l}^{y})}{\Delta x_k}-\frac{( S_{k,l+1}^{x}- S_{k,l}^{x})}{\Delta y_l} \right ),
\end{equation}
where $\Delta x_k$ and $\Delta y_l$ are  components of interparticle vector connecting nn spins and the sum is over indexes that characterize lattice coordinates. It can be demonstrated that $\rho$ is maximum for a perfect vortex configuration with  a value that depends on the tube radius and length.   
In analogy to the susceptibility associated to magnetization, we also define an effective susceptibility to characterize transitions associated with the appearance of states with finite vorticity:
\begin{equation}
  \chi_{\textsf{eff}}=\frac{\langle (\rho^z)^2\rangle- \langle \rho^z \rangle^2}{k_B T N} \ .
\end{equation}

\begin{figure}[tbp]
\centering
\includegraphics[width=0.8\columnwidth]{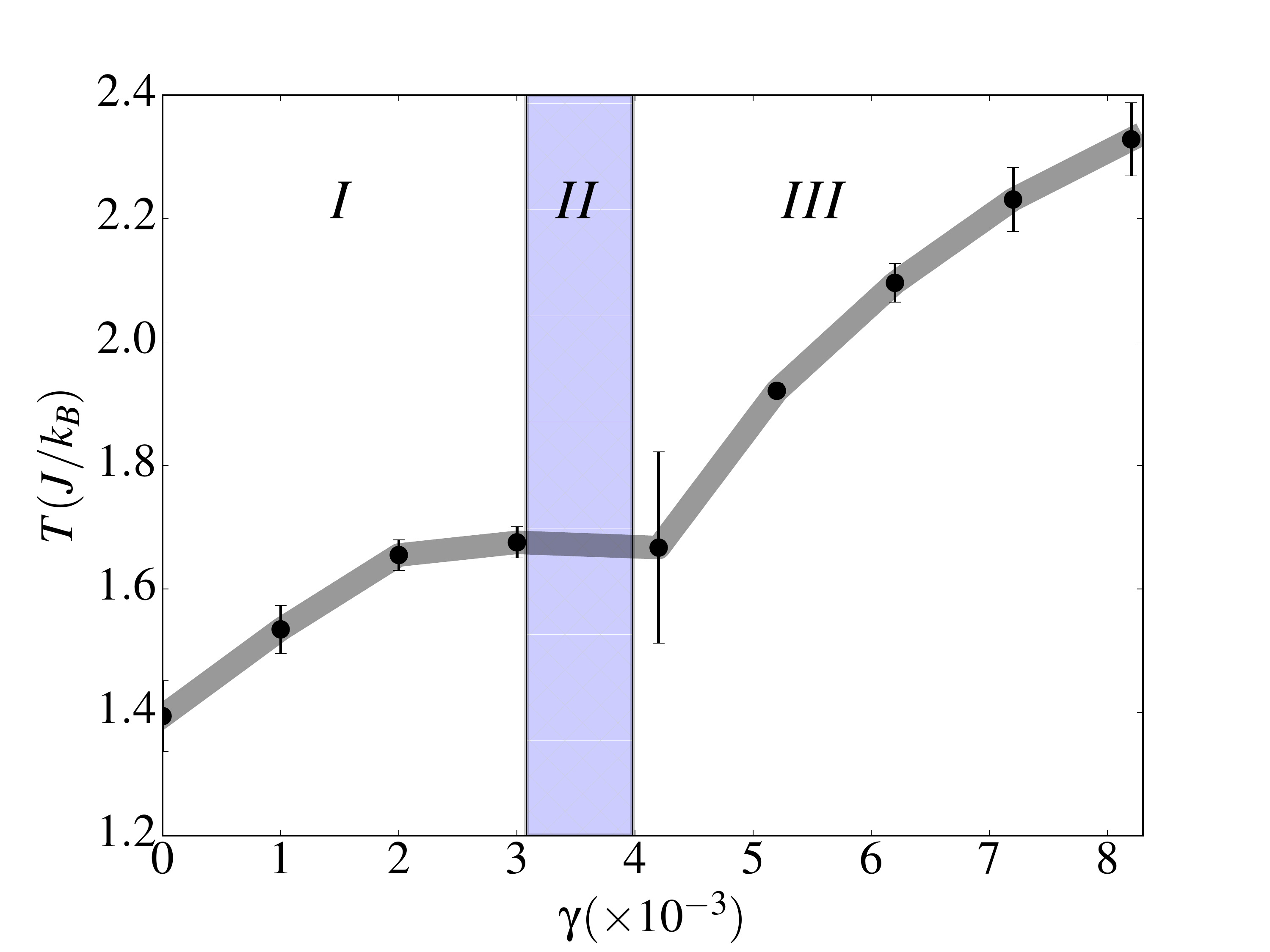}
 \caption{Transition temperature as function of $\gamma$ for a (8,15) tube. Region I stands for FM-PM transitions whereas region III stands for V-PM transitions. In region II, H-PM transitions are present and the system is highly metastable. Error bars correspond to averages over $5$ independent runs.}
\label{Fig_Critical-temperature}
\end{figure}

As shown in Fig. \ref{Fig_Thermaldependence2}(a), in the regime $\gamma\ge 0.04$, the normalized vorticity behaves as expected for an order parameter. It attains a finite value at low temperature that saturates to a value that corresponds to a perfect vortex (V) state having the spins pointing tangent to the tube surface and perpendicular to the tube axis. 
The transition temperatures $T_c$ for FM-PM and V-PM transitions were extracted by averaging the locations of the maxima  of the corresponding susceptibilities [see Fig. \ref{Fig_Thermaldependence2}(b)] and specific heat  and their dependencies on  $\gamma$ are shown  in Fig.~\ref{Fig_Critical-temperature}. Results indicate that $T_c$ increases with $\gamma$ for both transitions as the dipolar interaction becomes stronger. However, for the V-PM transition, $T_c$ seems to saturate when the dipolar energy dominates over exchange, in agreement also with finite size scaling theory \cite{Fisher} since a $\gamma$ increase can also be related to a size increase.

For intermediate values $0.03<\gamma<0.04$, the magnetization begins to exhibit a noisy behavior with big error bars [see Fig. \ref{Fig_Thermaldependence}(d)], although characterized by non-vanishing values, despite of the several configurational averages performed. Concomitantly, the vorticity  [see Fig. \ref{Fig_Thermaldependence2}(a)] behaves less noisy and acquires finite values at low temperatures, which means that certain degree of circulation of the spins around the tube surface emerges. Thus, the low temperature configurations are characterized by helical (H) states \cite{Salinas_jsnm2012} having spins pointing in the plane tangent to the tube surface as for the V states, but now with some misalignment relative to the tube axis. 

\begin{figure}[tbhp]
  \centering
    \includegraphics[width=0.9\columnwidth]{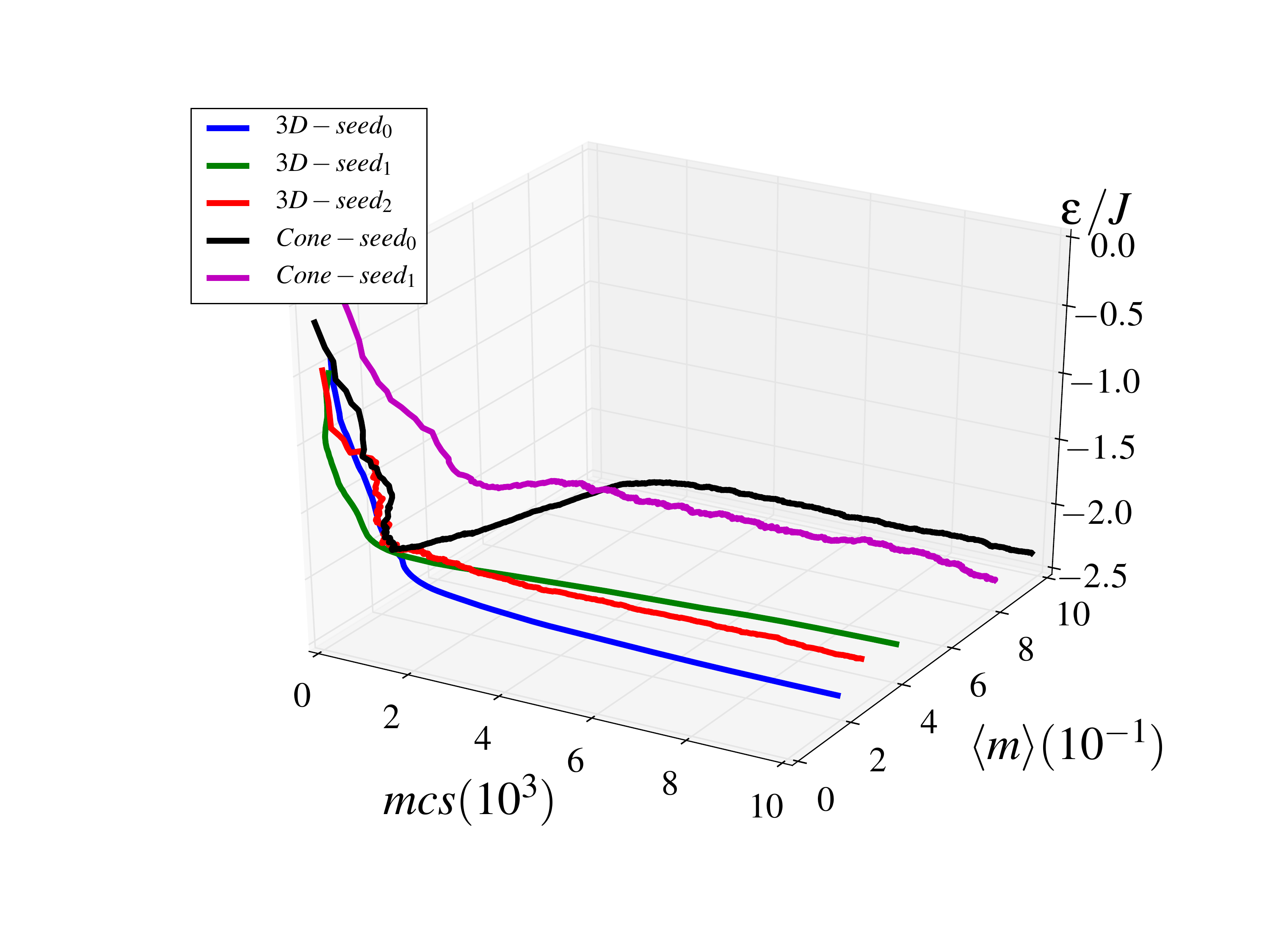}
    \caption{Energy dependence with both the number of MC steps and the magnetization per site for an intermediate $\gamma=0.035$ value where a H-PM transition is expected. Lines referred as 3D in the legend correspond to the cooling protocol (1) mentioned in the main text, whereas those referred as Cone correspond to protocol (2).
		Results, revealing a high metastability, were obtained by using several initial random seeds (different colors) and low temperature $T=0.1\ ,J/k_B$.}
  \label{Fig_meta}
\end{figure}
For this intermediate range of $\gamma$, it was not possible to infer precisely a H-PM transition temperature due to the high degree of metastability. The reason for this can be clearly noticed in Fig. \ref{Fig_meta}, where several long runs using the previously mentioned protocols $1$ and $2$ led to different low temperature states through different paths along the energy landscape, a situation that does not happen for lower or higher $\gamma$ values. Therefore, in order to study the low temperature configurations, we ran additional simulations using simulated annealing protocol $3$ varying the temperature according to a $T= T_0 0.95^k$  dependence and sampling trial spin movements in a cone with progressive reduction of its aperture in order to ensure convergence to the lowest energy states.

\begin{figure}[htb]
\centering
\includegraphics[width=0.9\columnwidth]{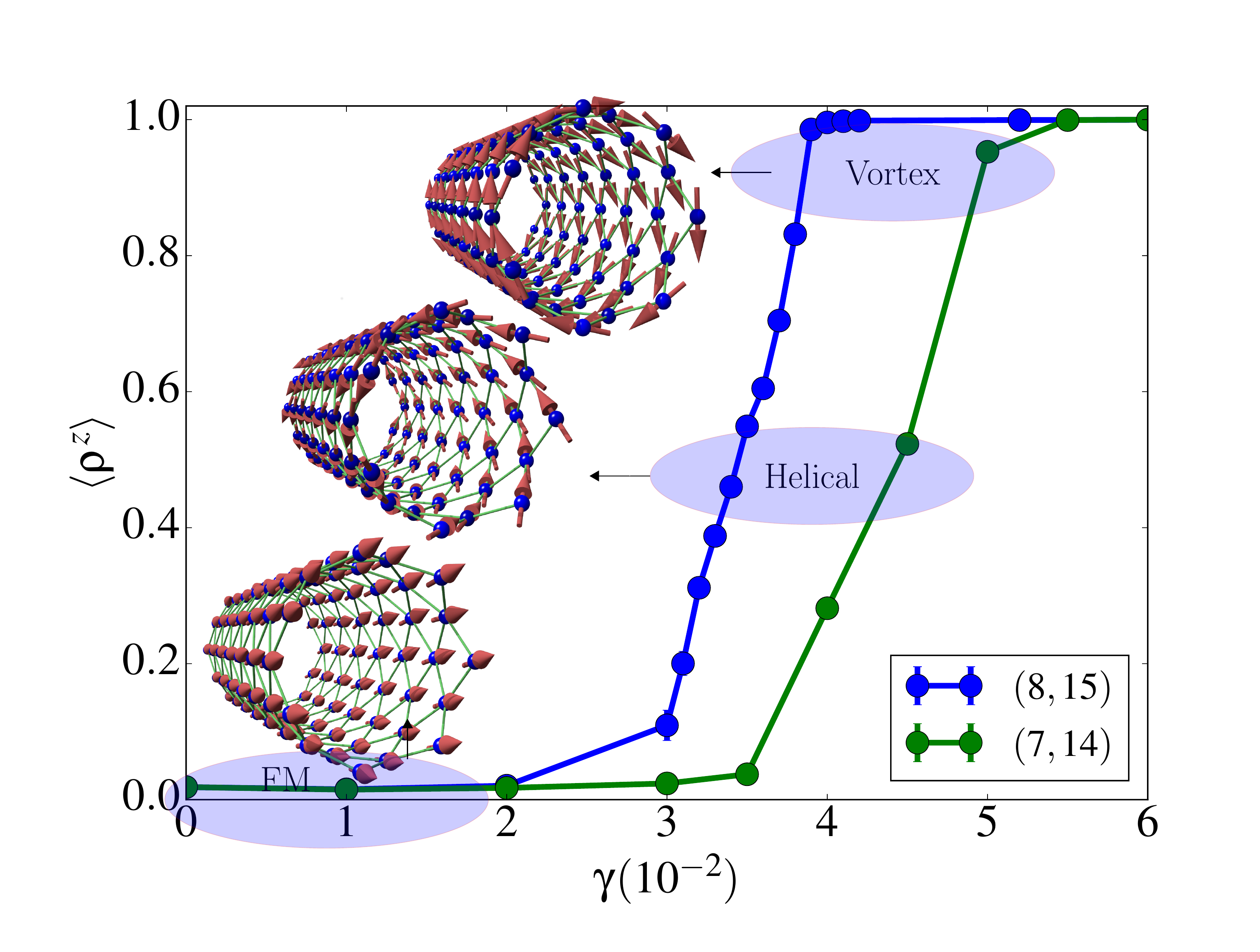}
\caption{Dependence of the normalized vorticity $\rho^z$ on $\gamma$ calculated for the low temperature configurations obtained using simulated annealing for tubes of size (8,15) and (7,14). The continuous lines serve to define a phase diagram separating the regions corresponding to to FM, Helical and Vortex states, whose corresponding magnetic configurations are shown in the insets for a (7,14) tube. Blue spheres correspond to lattice nodes and red arrows to spin directions.}
  \label{Figure_Vorticity}
\end{figure}

By using the vorticity as an order parameter, we were able to obtain the phase diagram shown in Fig. \ref{Figure_Vorticity} for two different tube sizes. The phase diagram evidences an increase in the vorticity of the system as the strength of the long range dipolar interaction increases. For both sizes, pure vortex states are stable at the highest considered values of $\gamma$, and this is the reason why $\gamma$ values greater than $0.06$ were not explored. 
Finally, we point out that a finite-size effect can be observed on the phase diagram. With a slight variation of the tube radius and length from (8,15) to (7,14), the onset of stable vortex states takes place at smaller $\gamma$. We will further elaborate on this point in a subsequent subsection.

\subsection*{Ground state characterization}
\label{Ground State_Sec}

\begin{figure}[tbp]
\centering
  \includegraphics[width=\columnwidth]{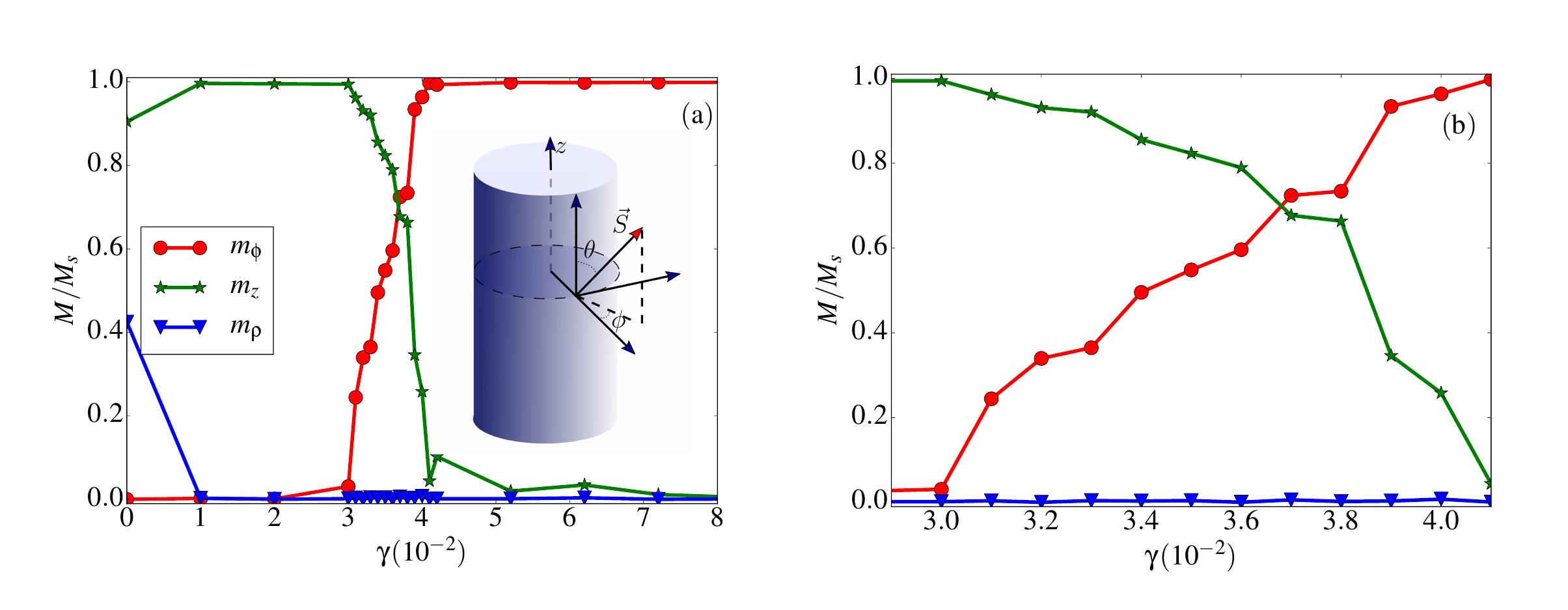}
  \caption{(a) Cylindrical components of the total magnetization per site of the low temperature magnetic configurations obtained by simulated annealing as a function of the $\gamma$ parameter. (b) Zoom of the intermediate region where transition from ferromagnetic alignment to vortex states takes place. The inset in (a) shows the coordinates used to describe the magnetic configurations.}
  \label{Figure_Translongmagn}
\end{figure}

In order to better characterize magnetic ground state configurations, we have expressed the total magnetization of the tubes in cylindrical coordinates (azimuthal  $m_\phi$  tangential to the tube surface, radial $m_\rho$ and longitudinal $m_z$). By projecting the spin vectors along a local reference frame centered on the spin positions $\vec{r_i}=(R\cos\varphi_i,R\sin\varphi_i,z_i)$  (see inset in Fig. \ref{Figure_Translongmagn}) 
the spin vector components can be written as:
\begin{eqnarray}
{S_{\rho}}_i=\sin\theta_i\cos\phi_i \nonumber\\
{S_{\phi}}_i=\sin\theta_i\sin\phi_i\nonumber\\
{S_z}_i={S_z}_i=\cos\theta_i \  ,
\label{Eq_Spherical}
\end{eqnarray}
so the vortex and helical states formed at intermediate $\gamma$ values can be better described. The  magnetization components of the  configurations obtained at the lowest temperature after simulated annealing are plotted in Fig. \ref{Figure_Translongmagn} for different values of $\gamma$. 
For $\gamma> 0.04$,  the configurations present vanishing radial and longitudinal magnetizations (i.e. $m_{\rho}\approx 0$ and $m_{z}\approx 0$) and a tangential component $m_{\phi}\approx 1$, characteristic of a V state perpendicular to the $z$ axis. In contrast,  for $\gamma< 0.03$  the obtained configurations have vanishing radial and tangential magnetizations (i.e. $m_{\rho}\approx 0$ and $m_{\phi}\approx 0$) and a longitudinal component $m_{z}\approx 1$, characteristic of FM alignment along the tube axis.
However, in the region $0.03<\gamma<0.04$ [see inset in Fig. \ref{Figure_Translongmagn}(a)], configurations have only vanishing radial magnetization, while both $m_{z}$ and $m_{\phi}$ have non-vanishing values that decrease (increase) with increasing $\gamma$.This indicates that H states can be seen as a combination between V and FM alignment that originates from the competition between the exchange interaction trying to align spins along a single direction and dipolar interactions that favor circularity with spins tending to point tangent to the surface of the tube to reduce magnetostatic energy. 

These helical states can be better characterized by plotting the angle $\theta$ averaged over all the $\theta_i$ of spins in every layer, i.e. for each $z$ value along the length $L$ of the tube, as shown in Fig. \ref{Figure_Averageangle}.  An alternative representation is given in Supplementary Fig. S2, where all the spins of the tube are represented by dots on a unit sphere according to their orientation and are given a color that depends on the value of $\gamma$ of the considered configuration. 
\begin{figure}[htb]
  \centering
    \includegraphics[width=0.8\columnwidth]{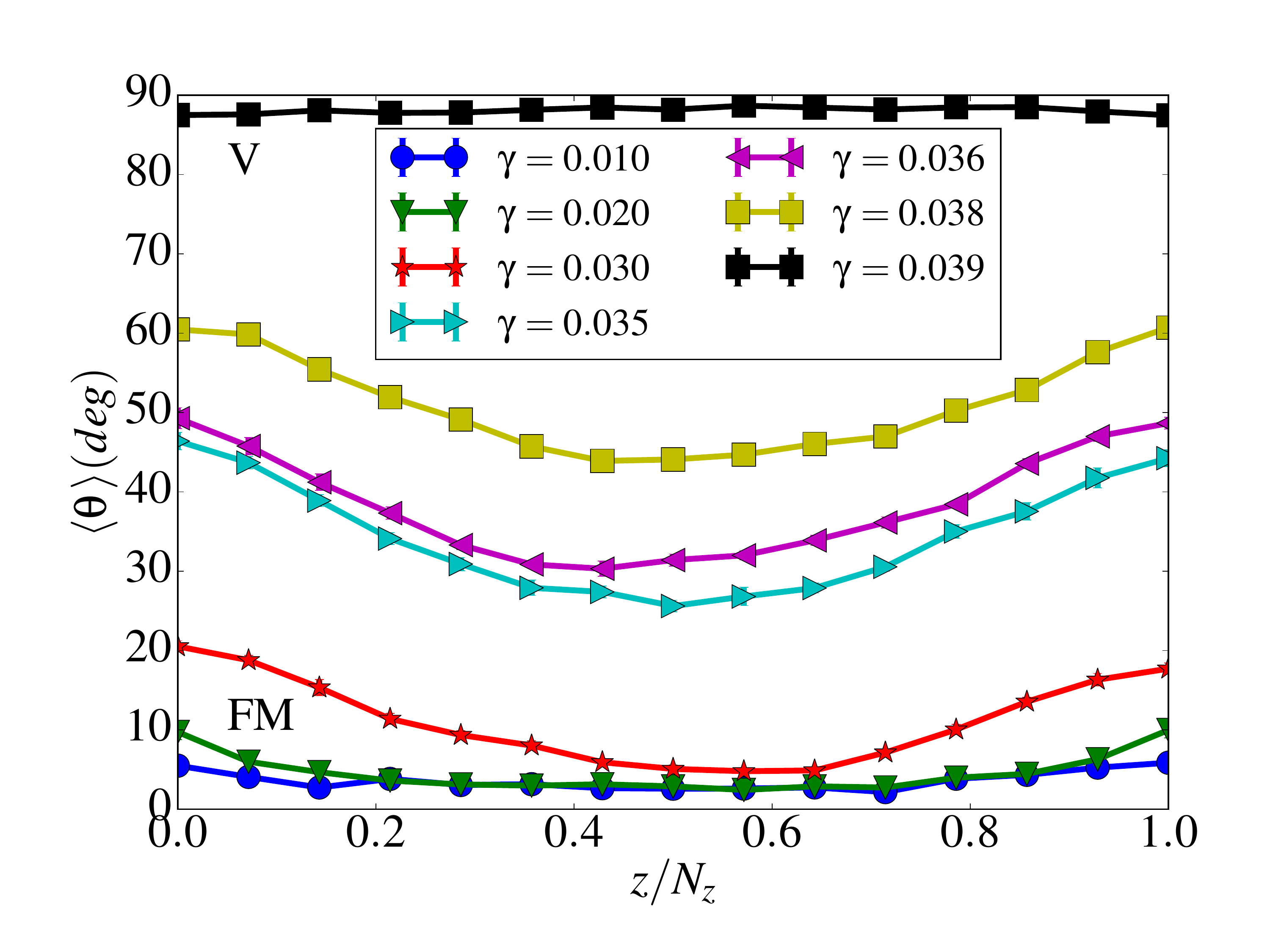}
   \caption{Height profile of the average polar angle $\left<\theta\right>$ for a $(8,15)$ tube. Results for intermediate values of $\gamma$ are given in the left panel and while in the right panel results are shown for the highest $\gamma$ values considered in the simulations. In the right panel, $\left<\theta\right>\approx$ 30$^{\circ}$ stands for a FM alignment whereas values of $\left<\theta\right>$ close to 90$^{\circ}$ correspond to pure vortex states.}
 \label{Figure_Averageangle}
\end{figure}

As it can be seen in  Fig. \ref{Figure_Averageangle}, the angle $\langle \theta\rangle$ is not uniform along the tube length and tends to increase towards the tube ends for all the computed $\gamma$ values. 
Only for $\gamma=0$ the vorticity is strictly zero, and $\langle \theta \rangle$ is constant along the tube. For $\gamma \lesssim 0.3$, spins are still mostly aligned along some common $\langle \theta \rangle $ angle which deviates from $0$ (FM states) with increasing dipolar interactions , while for $\gamma \gtrsim 0.4$, all configurations average around $\langle \theta \rangle \approx \pi/2$. 
In the unit sphere representation of Supplementary Fig. S2, FM states appear as a groups of dots localized around the poles, while V states are revealed in the form of groups distinguishable along the equator of the sphere (note that the amount of groups equals twice the number of spins per layer due to geometrical considerations and the discreteness of the lattice). 

However, for intermediate values in the region $0.03<\gamma< 0.04$ where H states arise, profiles are not longer flat evidencing different spin configurations at the edges and the inner part of the tube. More concretely, $\left<\theta\right>$ is lower in the middle part of the tube and greater at the edges. This means that the degree of vorticity is higher at the ends than at the central part of the tube. This behavior can be more clearly observed for longer tubes, as we will show in the next section. Due to the competition between exchange and dipolar energies, helical states transform onto vortex states as $\gamma$ increases, and this change occurs first on the rings near the tube edges.

\subsection*{Size dependence and phase diagram}
\label{Size_dep_Sec}
\begin{figure}[tbp]
  \centering
  \includegraphics[width=\columnwidth]{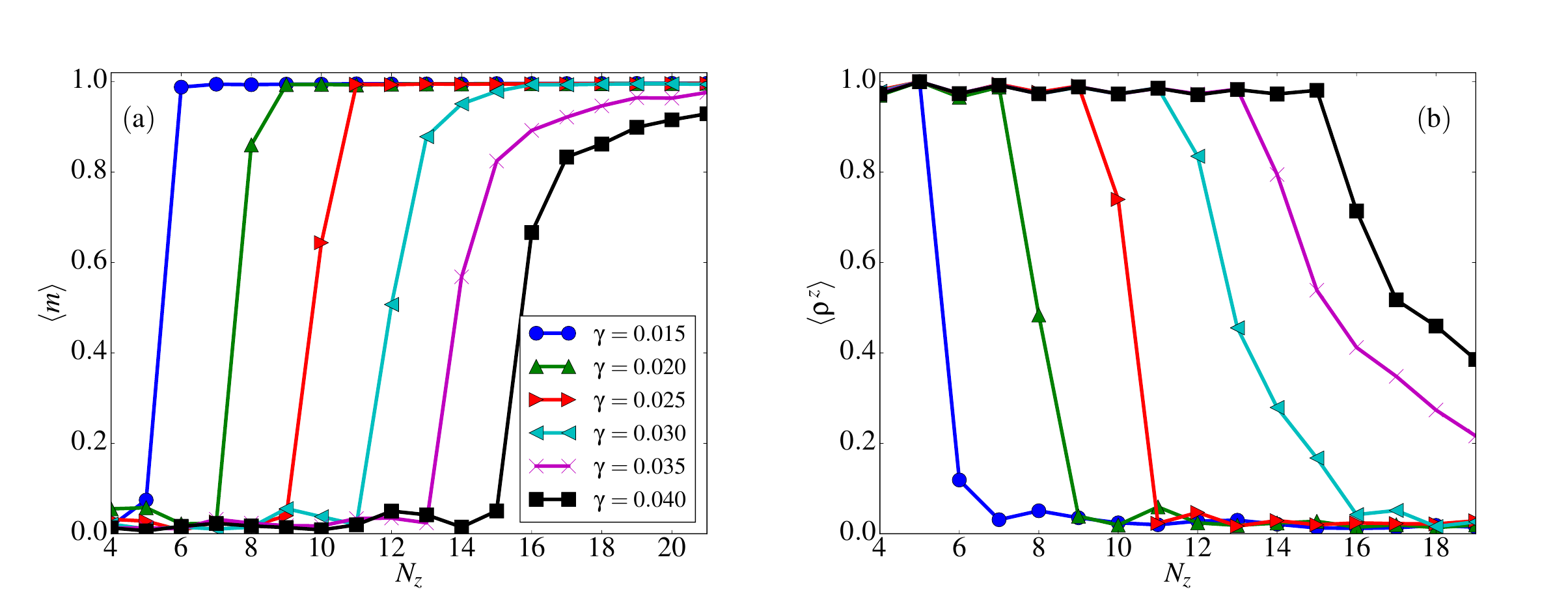}
  \caption{Magnetization (a) and $z-$component of the vorticity (b) of the equilibrium configurations as a function of the number of layers $N_z$ of a tube with $N=8$ for $\gamma \in [0.015, 0.040]$ as indicated in the legend.}
  \label{Fig_altura_mag_vor}
\end{figure}

In the previous subsections, we focused on determining the low energy configurations as a function of $\gamma$ for a fixed tube radius and length, exemplified by a (8,15) tube. Now, we carry out  a systematic study of the low temperature configurations for different values of $N_z$, $N$ at different fixed interaction strengths $\gamma$. 
For this purpose, we start first by varying the tube length while keeping the number of spins per layer at $N=8$. Using simulated annealing, we have obtained the low temperature magnetization and  vorticity of the low temperature configurations, as shown in Fig. \ref{Fig_altura_mag_vor} for a number of layers $N_z$ in the range from $4$ to $25$ and $0.015\leq\gamma \leq 0.045$.

Length driven transitions from FM to V states are identified by the $N_z$ for which $\langle m \rangle \approx 1$ and $\langle \rho^z \rangle \approx 0$  and transitions from FM or H to V states by $\langle m \rangle \approx 0 $ and $\langle \rho^z\rangle \approx 1$ , while intermediate values of these parameters characterize regions where H states are favored. As observed in Fig. \ref{Fig_altura_mag_vor}, transitions are sharp for $\gamma<0.03$, reflecting the fact that, in this range, exchange interactions dominate and states with almost uniform alignment along the tube axis are favored. Only for tubes with low aspect ratio, formation of V states is observed. Starting at values $\gamma>0.03$,  transitions between $\rho^z=0$ and $1$ become smoother and the region in between indicates the interval of $N_z$ for which H states appear. With increasing values of $\gamma \geq 0.04$, V states prevail up to higher tube lengths, until a point where formation of FM states is not favored even for the longest simulated tubes.

As an example of what happens for longer and wider nanotubes, we show representative magnetization height profiles for $\gamma=0.05$ in Fig. \ref{ferro-vor01}. Snapshots of the spin configurations are reported in Supplementary Fig. S5. As it can be observed, for the thinner $N=8$ tubes, increasing the length leads to an increase of the central part with FM aligned spins at the expenses of the decrease of the vortices at the tube ends and also of the width of the domain walls connecting the vortices to the central FM region.

For wider tubes ($N= 16$), a perfect V configuration is obtained for short tubes but this transforms into a H state when the length increases. However, even for a tube with aspect ratio $1:4$, only the central layer of spins is oriented along the tube axis. Much longer tubes have to be considered in this case for the observation of an extended FM central region, due to the strong in-plane dipolar fields generated by the vortices formed at the tube ends.  

\begin{figure}[tbp]
  \begin{center}
    \includegraphics[width=0.8\columnwidth]{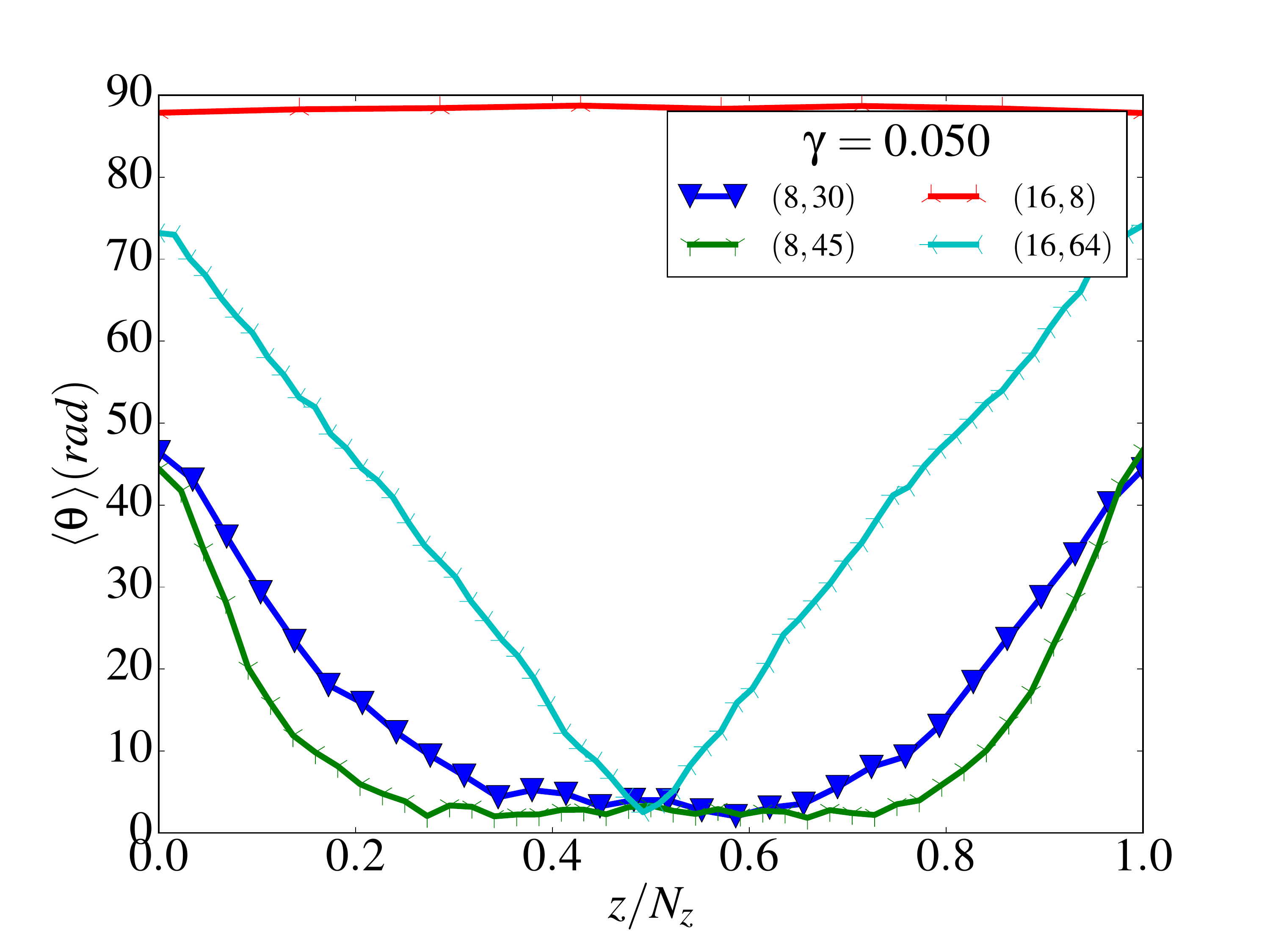}
  \end{center}
  \caption{Typical longitudinal profile of the average polar angle $\left<\theta\right>$ for $\gamma = 0.05$ nanotubes with different radii and lengths as indicated in the legend. FM alignment in the central part of the tube and a trend to form vortices at the edges is observed as the tube length is increased.} 
  \label{ferro-vor01}
\end{figure}

\begin{figure}[htb]
\centering
\includegraphics[width=0.8\columnwidth]{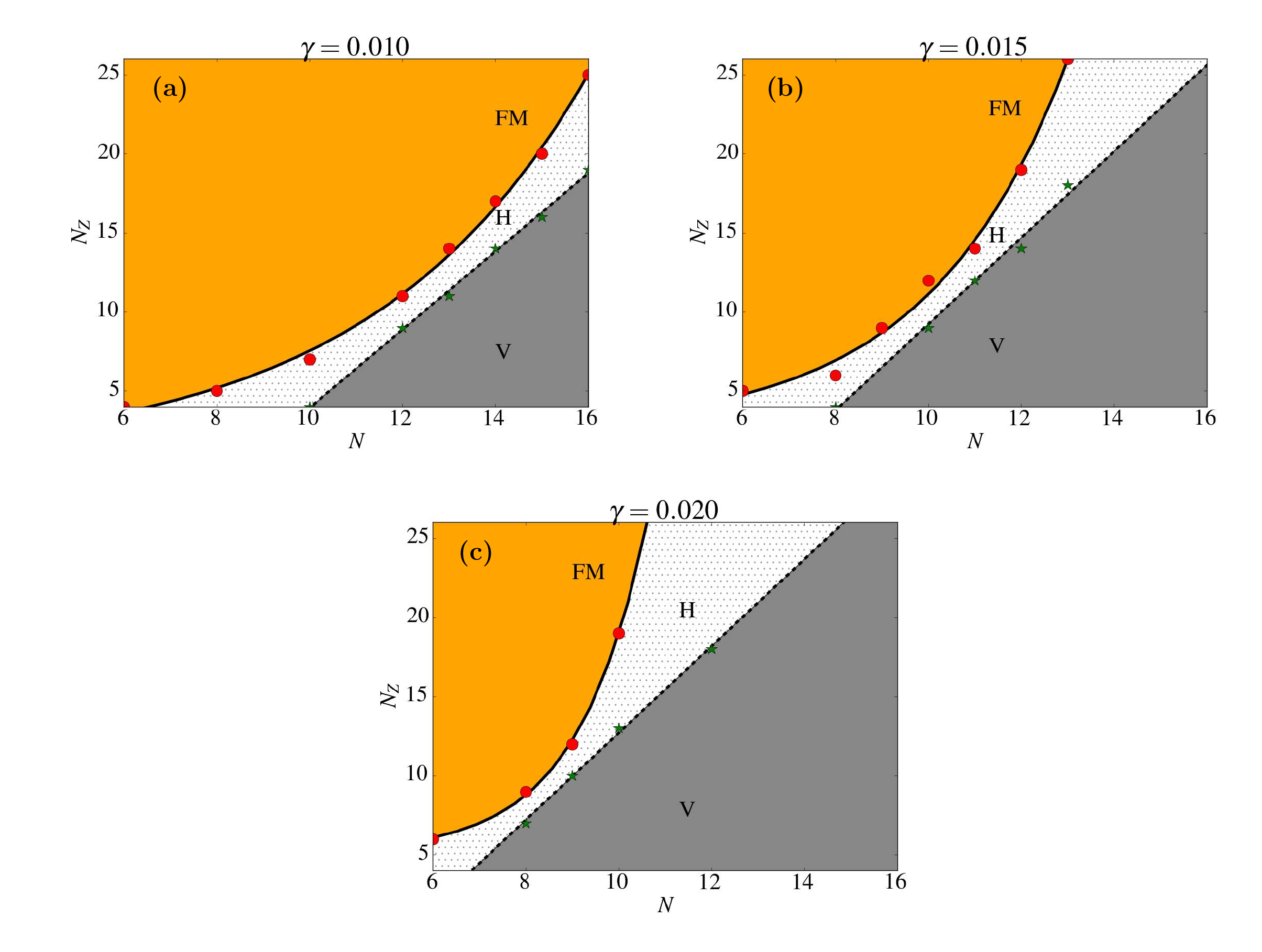}
\caption{Configuration phase diagrams for tubes with different aspect ratios $(N/Nz)$ and three values of  $\gamma=0.01, 0.015, 0,02$.  Red circles identify FM-H transitions and green stars V-H transitions. Continuous lines are fits of the transition curves to the laws described in the text.}
\label{Fig_diag_phases}
\end{figure}

We have extended the calculations to additional tube radii N=$10, \dots, 16$, using the procedure described previously in order to identify the $N, N_z$ at which FM-H and H-V transitions are observed. We have obtained the phase diagrams in the  ($N,N_z$)  plane shown in Fig. \ref{Fig_diag_phases} for three values of $\gamma$. We realize that the V-H critical curve can be fitted to a linear law $N_z=A+B N$ (dotted line) with $A$ and $B$ values given in Table \ref{Tab_1}. 
As observed, when increasing $\gamma$, the slope of this line increases and the $N$ axis intercept tends to zero, meaning that, even though if exchange is dominant, V states can be stabilized for low aspect ratio tubes and big enough radii. 
However, the line H-FM curve exhibits a more pronounced change with the radius of the tube. We have found that this line can be fitted to an exponential dependence of the form $N_z=N_0+A_2\exp(N/B_2)$ with $A_2$ and $B_2$ values given in Table \ref{Tab_1}.  Comparing the results for different $\gamma$, we see that there is always a narrower region for formation of H states for tubes having length equal to radius ($\sim 12, 10, 8$ for $\gamma= 0.01, 0.015, 0.02$) and that is shifted to lower $N$ with increasing $\gamma$.
  
 \begin{table}[th]
  \begin{center}
    \begin{tabular}{| c || c | c || c | c | c | }
      \hline
      $\gamma$ &   $A$         &  $B$       &  $N_0$       &   $A_2$         &  $B_2$\\ \hline
      $0.01$  &$ -21.0 $  &  $2.50$ & $0.98$ &$ 0.71 $  &  $4.54$ \\ 
      $0.015$  &$-16.3$  &  $2.54$  & $1.45$ &$ 0.54$  &  $3.44$ \\ 
      $0.02$  &$ -14.74 $  &  $2.74$  & $5.14$ &$ 0.017 $  &  $1.49$\\ \hline
\end{tabular}
\end{center}
 \caption{Results of the fits of the curves separating the V-H and FM-H configurations shown in Fig. \ref{Fig_diag_phases}. Values for $A$ and $B$ are the fitted parameters to a linear law for the V-H curve and $A_2$ and $B_2$ are the fitted parameter to an exponential law for the FM-H curve.}
 \label{Tab_1}
 \end{table}

\subsection*{Comparison to analytical calculations}
\label{Analy_Sec}
Given the narrow region of $\gamma$ values for which H configurations with complex magnetic order are stabilized, next we will study their metastability by comparing their total energy to that of simpler configurations that interpolate continuously between  FM and V states. 

\begin{figure}[tbp]
  \centering
  \includegraphics[width=\columnwidth]{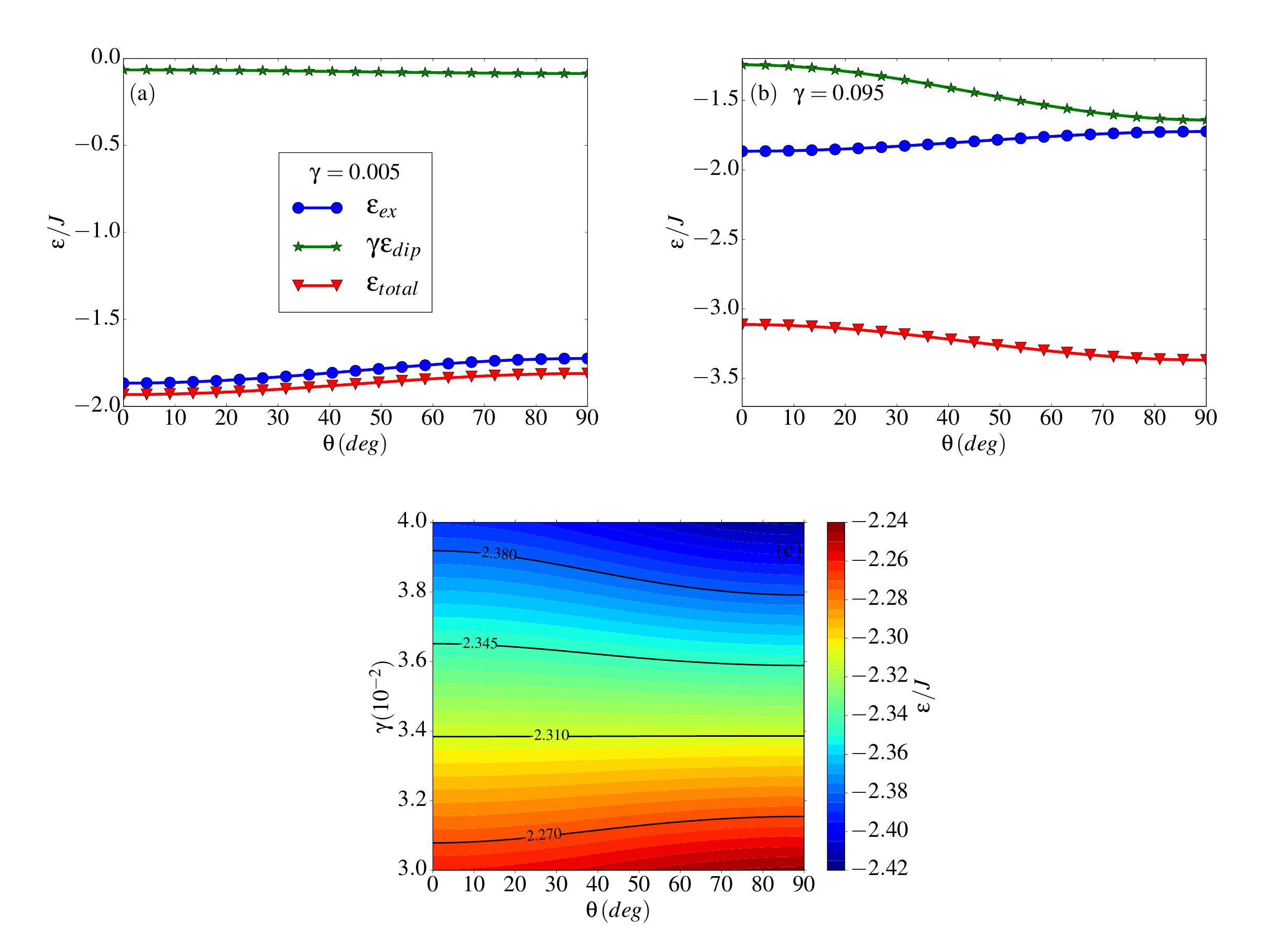}
  \caption{Dependence of exchange, dipolar and total energy per spin on the polar angle $\theta$ for a configuration with spins tangential to the tube surface all forming an angle $\theta$ with respect to the tube axis. Results are shown for a (8,15) nanotube. Panel (a) is for $\gamma = 0.005$ and panel (b) is for $\gamma = 0.095$. Panel (c) displays a contour plot of the dependence of the total energy on $\theta $ for  $\gamma$ values close to the formation of H states, where isoenergetic curves are plotted with black lines.}
 \label{Fig_EvsTheta}
\end{figure}

\begin{figure}[tbhp]
  \centering
  \includegraphics[width=0.8\columnwidth]{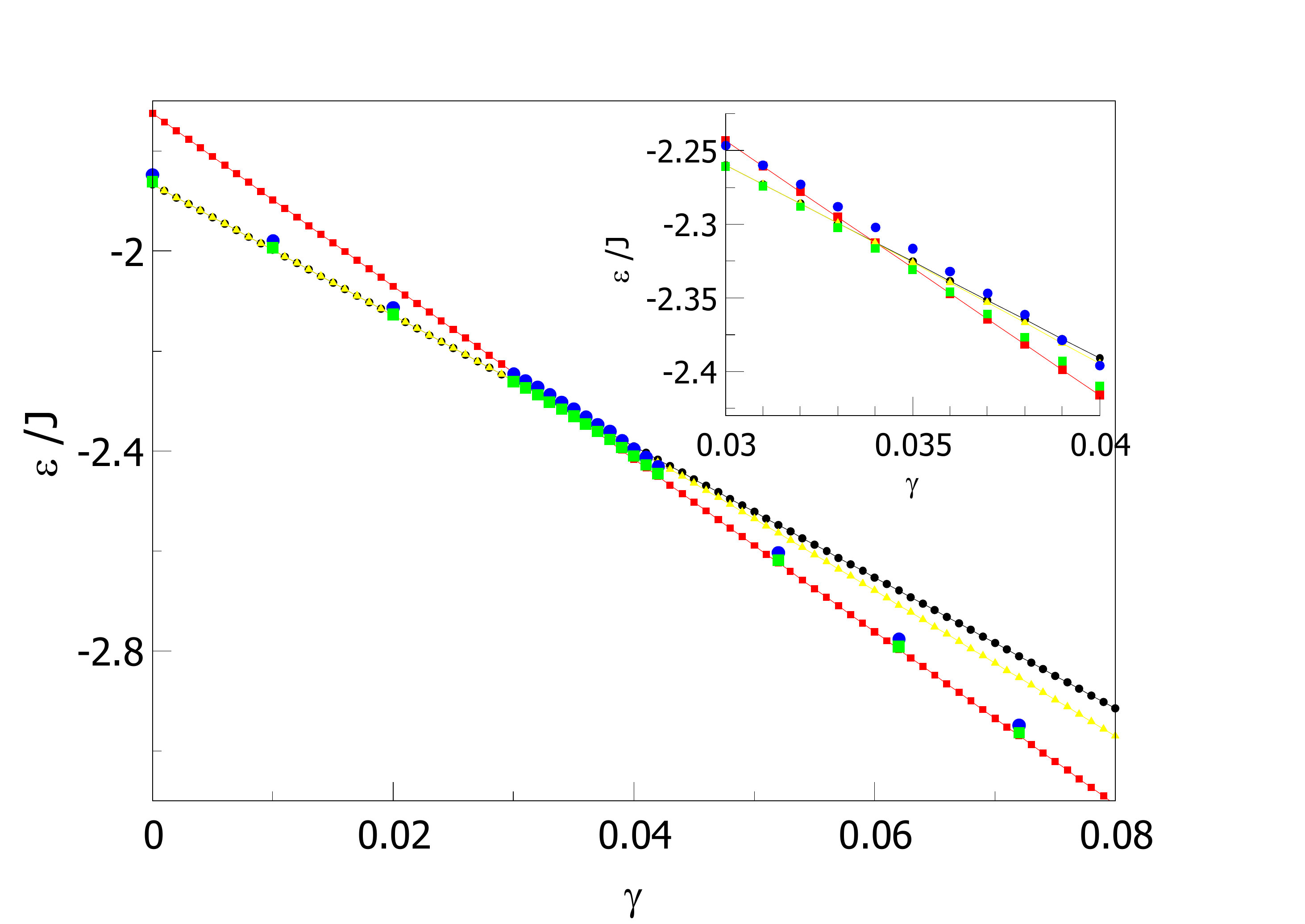}
  \caption{Energies of ground state configurations of a (8,15) tube obtained from the simulations following an annealing (green squares) or thermalization (blue dots) procedures are compared to those of configurations having spins oriented at $\theta=0$ (FM state, black circles) or $\theta=\pi/2$ (V state, red squares) tangentially to the tube surface. Yellow triangles correspond to analytical results for the mixed states with minimum energy as described in the text. Inset: zoom around the range $0.03<\gamma <0.04$.}
    \label{Fig_EnergiesGamma}
\end{figure}

\subsubsection*{Quasi-uniform states}
\label{Sec_Exact_Unif}
As a first case, we have considered configurations with the spins pointing along the local plane tangent to the tube surface and all oriented along a direction characterized by a polar angle $\theta$ with respect to the tube axis.
We start computing the exchange energy per spin in $J$ units (i.e. total exchange energy divided by $N_{tot}=NN_z$) , which can be shown to be given by (see Supplementary Information for details):  
\begin{equation}
\label{ec:exchange-energy}
  \epsilon_{ex}=-\frac{z}{2}\left( 1-\frac{1}{N_z}\right)\left[1-\sin^2\theta(1-\cos(\varphi_N/2)) \right] \ ,
\end{equation}
where $z$ is the coordination number ($z=4$ in our case)$,  \varphi_N=2\pi/N$ is the angle subtending two consecutive spins on the same layer containing $N$ atoms, whereas $\varphi_N/2$ is the basal projection angle between two n.n. interacting spins belonging to consecutive layers according to the zig-zag geometry. Thus, exact values can be obtained for the FM or V configurations by setting $\theta =0, \pi/2$ in Eq.  \ref{ec:exchange-energy}, while intermediate values of $\theta$ would correspond to H states.

Notice that the energy depends on the tube radius (or number of spins per ring $N$) through $\varphi_N$ and also on $N_z$ due to  the finite tube length. 
In the long tube limit  $N_z\rightarrow\infty$, it becomes $\epsilon_{ex}=-(z/2)$ for the FM state and $\epsilon_{ex}=-(z/2)\cos(\varphi_N/2)$ for the vortex state, as expected.  Due to the $\sin^2\theta$ dependence,  $\epsilon_{ex}$ increases progressively as the system evolves from FM to V states passing through intermediate H states.  The variation of this energy with $\theta$ is shown in Fig. \ref{Fig_EvsTheta} in blue circles for a $(8,15)$ tube.

The dipolar energy can also be computed using analytical expressions, although their derivation is more complex due to the long-range character of the interactions and the fact that contributions from spins in the same ring and from different (odd or even) rings  have to be considered separately (see Supplementary Information for the details). As shown in Fig.  \ref{Fig_EvsTheta} (green symbols), $\epsilon_{dip}$ is minimized for $\theta=\pi/2$ and increases when the magnetic moments progressively align along the tube axis. Therefore, as a consequence of the competition between exchange and dipolar energies, the state with minimum energy will depend on the value of $\gamma$.  This can be clearly observed by looking at the total energy plotted in Fig. \ref{Fig_EvsTheta}(a), (b) for two different values of $\gamma$: for small $\gamma$, the FM state with $\theta=0$ is favored, while for large $\gamma$ the V state has the lowest energy.

By looking at the evolution of the total energy when going from low to high  $\gamma$, it is clear that there must be an intermediate $\gamma$ for which $\epsilon_{tot}$ does not depend on the polar angle $\theta$. This value can be found by equating the total energies per spin for $\theta=0,\pi/2$, which results in
\begin{equation}
\label{ec:gamma-star}
 \gamma^\star=\frac{\epsilon_{ex}(\pi/2)-\epsilon_{ex}(0)}{ \epsilon_{dip}(0)-\epsilon_{dip}(\pi/2)}\ \ .
\end{equation}
Inserting the values for the (8,15) tube (see Supplementary Information), this gives $\gamma^\star=0.034$ with a value for the energy $\epsilon_{tot}/J= -2.31$ which lies precisely at the center of the region of $\gamma$ where H states are obtained in the simulations.  We notice that, as can be seen in Fig. \ref{Fig_EvsTheta}(c),  for $\gamma$ close to this critical value, configurations with different $\theta$ have energies that differ in less than $1 \%$ and, as a consequence, they are quasi-degenerated.  Therefore, within this range of $\gamma$'s, it is very difficult to ascertain if the configurations obtained from simulation are the true ground states, even if the annealing protocol is used.

\begin{figure}[tbp]
  \centering
  \includegraphics[width=0.8\columnwidth]{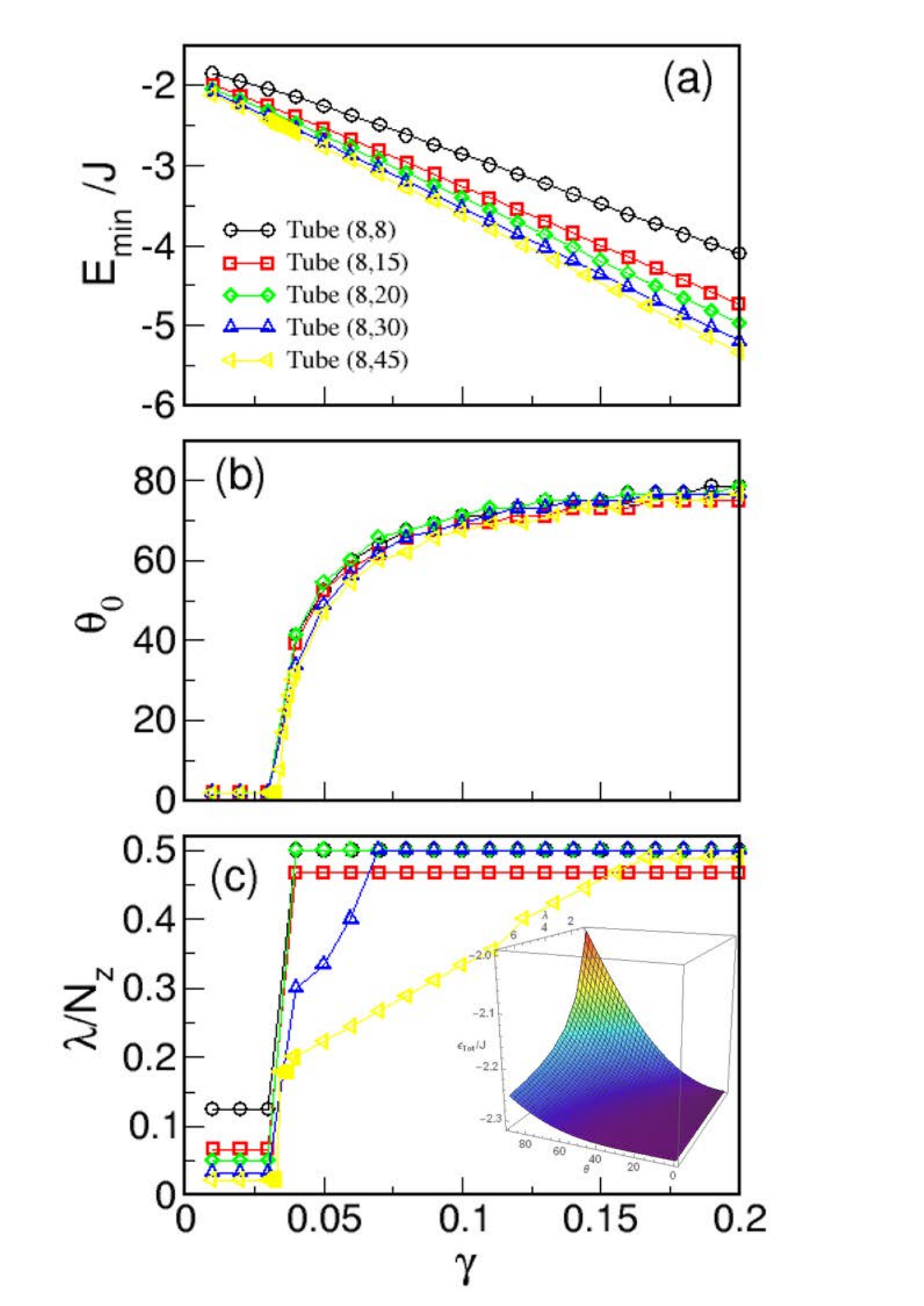}
  \caption{Results of energy numerical minimization of analytical mixed states characterized by a FM central region of length $\lambda$ and angle $\theta_0$ at the tube ends. 
	Dependence of (a) $E_{min}$, (b) $\theta_0$ and (c) $\lambda$, and on $\gamma$ is shown for tubes with a fixed radius $N= 8$ and different number of layers $N_z=  8, 15, 20, 30, 45$. The inset in panel (c) displays the energy surface for the $(8,15)$ tube with $\gamma=\gamma^\star= 0.034$.}
  \label{Fig_dvxth0_minHe}
\end{figure}

In fact, comparing  the energies of the states obtained from the simulated annealing process with those of the states with uniform $\theta$ (see Fig. \ref{Fig_EnergiesGamma}), we find that for $\gamma< \gamma^\star$ simulation results have exactly the same energy as the FM states whereas, for $\gamma>\gamma^\star$, configurations obtained through simulated annealing have slightly higher energies than V states.
Notice also that, in the same figure, we can see that configurations obtained after annealing (green squares) have indeed lower energies than those obtained after the usual thermalization process (blue circles) used to compute thermodynamic averages, as anticipated in the previous section.

Finally, we have computed the dependence of $\gamma^\star$ on $N_z$ as shown in Supplementary Fig. S3 for different tube radii. We observe that the region of $\gamma$ for which quasi-degenerated configurations are expected, is displaced to higher values of $\gamma$ as the tube length increases and that, for a fixed tube length, $\gamma^\star$ decreases when increasing the tube radius. This serves also as an indication for the expected location of the region favoring 
 H states and it is in agreement with what is found on the phase diagrams presented in  Fig. \ref{Fig_diag_phases}.

\subsubsection*{Mixed states}
\label{Sec_Exact_Mixed}

The low temperature H configurations that we have obtained by simulations present some non-uniformities at the tube ends that depart from the states with uniform $\theta$ considered in the previous subsection. Therefore, in an effort to find a better approximate analytical description for the equilibrium magnetization patterns, we will consider now the energy of configurations for which a central region (of length $l-2\lambda$) of the nanotube with a uniform FM state is connected by a pair of incomplete vortex walls that extend towards the two ends of the tube, where the spins form an angle $\theta_0$ with respect to the tube axis \cite{Landeros_prb2009}. An example of such a configuration is displayed in Supplementary Fig. S4.

For this kind of configurations, no analytic expressions can be obtained for the energies. Therefore, for a given tube with length $L$ and radius $R$, we have calculated numerically energies for a selection of values of $\theta_0\in [0,\pi/2]$ and $\lambda\in [0,L/2]$ and located the minima of the $E(\theta_0,\lambda)$ surface for a range of $\gamma$ values. 
The results of this minimization procedure are shown in Fig. \ref{Fig_dvxth0_minHe}, where the $\gamma$ dependence of  $\theta_0$, $\lambda$ minimizing the energy [panels (b) and (c)], and the corresponding energies $E_{min}$ [panel (a)] are shown for tubes with $N= 8$ and different lengths. Similar results have been obtained for tubes with radi up to $N= 16$ (not shown).

First, let us stress the perfect linear dependence of  $E_{min}$ on $\gamma$ that indicates a linear scaling of the energy on this parameter when considering the values of $\theta_0$, $\lambda$ that minimize the energy. 
Second, regarding the angle at the ends of the tubes and the central region with uniform magnetization, we observe that, for $\gamma < \gamma^{\star}$, $\theta_0 \approx 0,\lambda\approx L$ and, therefore, FM configurations are obtained independently of the tube length. 
For $\gamma > \gamma^{\star}$, $\theta_0(\gamma)$ follows the same trend independently of the tube length, increasing progressively towards $90^{\circ}$ since for higher $\gamma$ the dipolar interaction favors flux closure at the ends. 
For the shortest considered tubes ($N_z\leq 15$), the vortex region extends to the central part of the tubes, whereas for long tubes [see for example the case $N_z= 45$ in Fig. \ref{Fig_dvxth0_minHe} (c) ] this region progressively expands as $\gamma$ is increased.

Finally, we have compared the energies of these optimized mixed states to those of the configurations from our MC simulations. As indicated by the yellow triangles in Fig. \ref{Fig_EnergiesGamma}, these states have lower energy than FM ones for $\gamma > \gamma^{\star}$ but clearly higher energies than those obtained by both simulation procedures, which are closer to the energies of perfect V states. 
This result is in contrast with the work of Landeros et al. \cite{Landeros_prb2009}, who obtained good agreement between simulations and optimized mixed states. However, they considered tubes with AA stacking of the spins instead of zig-zag stacking as in our case, and this may make a difference concerning magnetic properties. To confirm this, we have checked (not shown) that for AA tubes of the same size as ours, the energies of the optimized mixed states are indeed lower than perfect V states. This means that the spatial distribution of the spins on the surface of the nanotube has an influence on their magnetic order due to the anisotropic and the long-range character of the dipolar interaction. 

\section*{Discussion and Conclusions}
\label{Conclusions_Sec}

Using atomistic MC simulations, we have studied the ground state configurations that arise as a result of the competence between exchange and dipolar interactions in magnetic nananotubes of finite size, establishing radius-length phase diagrams for their stability for different $\gamma$'s. We have shown that, for a given radius and length, ferromagnetic (FM), vortex (V) or helical (H) states can be stabilized, depending on the value of $\gamma$. 
Our results are in agreement with experiments that have recently reported the observation of these states in real samples. We would like to emphasize that, although experimental studies have been conducted on nanotubes which have considerably bigger dimensions than those considered here, the application of scaling techniques \cite{Albuquerque_prl2002,Landeros_prb2005,Velasquez_prb2015}, would bring the range of $\gamma$ for the observation of the different states towards the order of magnitude of real material sizes .  

As mentioned previously, in our model Hamiltonian we did not include magnetocrystalline anisotropy as we were aiming to simulated magnetically soft materials. However, the results of previous studies \cite{Escrig_jmmm2007b,Chen_jap2010,Chen_jap2011,Lopez_jmmm2012} by means of micromagnetic simulation of nanotubes with finite width considering uniaxial or cubic anisotropy, can give a hint on the effect of its inclusion in our results.  
Therefore, we speculate that inclusion of uniaxial anisotropy along the tube axis in our model  would lead to an increase of the length of the central FM region and a reduction of the region $\lambda$ where vortices are formed at the tube ends. Probably, this would lead also to a reduction of the $\theta_0$ angle. Regarding the phase diagrams of Fig. \ref{Fig_diag_phases}, this would translate in displacement of the FM and V stability regions towards higher N values. However, it is difficult to guess in what way the transition lines between the three regions would be affected by the anisotropy. We are planning to address these issues in a forthcoming publication.

Furthermore, we would like to stress that our results are applicable to a wide variety of magnetic systems independently of the physical realization of the spins that reside on the nanotube surface. They could represent magnetic moments of atomic spins or of a magnetic cluster or nanoparticle as this would only change the values of $D$ and $J$. As shown in Supplementary Information, typical values of $D$ are in the range from $10^{-2}$ meV for atomic spins to $40$ meV for nanoparticles while $J$ vary in between $1$ and $10$ths of meV for most FM materials. Therefore, materials may be easily found with $\gamma$'s falling within the interval $10^{-2}-10^{-1}$ where we have found H configurations as ground states.

We have shown that a modified simulated annealing protocol has to be employed in order to be certain that the obtained configurations are indeed the true ground states, specially the ones showing helicoidal order.
In order to determine the interval of $\gamma$ values for which H states are the quasi-equilibrium configurations for a given radius and length of the nanotube, we have found crucial to introduce the so-called vorticity order parameter which clearly identifies the formation of states with zero magnetization but a well defined circularity or chirality. This has allowed us to establish the stability regions for their formation in radius-length phase diagrams, showing that, with increasing $\gamma$, the region of stability of H states becomes wider for higher aspect ratio tubes.   

In an effort to check the robustness of the helicoidal states, we have compared the energies obtained  by simulation to those of similar configurations that can be expressed analytically, such as quasi-uniform states with spins having a common angle with respect to the tube axis or mixed states formed by a FM ordered regions connected to vortex states at the tube ends. We have found that the states resulting from MC simulated annealing have indeed lower energies than any of the considered analytical ones  for the range of $\gamma$'s where the H states appear. However, the energy differences in this range are really small, indicating the high degree of metastability of the H states. This fact can be relevant for the nanotube reversal modes in hysteresis loops, which we will study in a future publication.


\begin{thebibliography}{10}
\expandafter\ifx\csname url\endcsname\relax
  \def\url#1{\texttt{#1}}\fi
\expandafter\ifx\csname urlprefix\endcsname\relax\def\urlprefix{URL }\fi
\expandafter\ifx\csname doiprefix\endcsname\relax\def\doiprefix{DOI }\fi
\providecommand{\bibinfo}[2]{#2}
\providecommand{\eprint}[2][]{\url{#2}}

\bibitem{Fernandez-Pacheco_natcomm2016}
\bibinfo{author}{Fernandez-Pacheco, A.} \emph{et~al.}
\newblock \bibinfo{journal}{\bibinfo{title}{Three-dimensional nanomagnetism}}.
\newblock {\emph{\JournalTitle{Nature Comm.}}} \textbf{\bibinfo{volume}{8}},
  \bibinfo{pages}{15756} (\bibinfo{year}{2016}).
\newblock \urlprefix\url{https://doi.org/10.1038/ncomms15756}.
\newblock \doiprefix 10.1038/ncomms15756.

\bibitem{Hertel_jpcm2016}
\bibinfo{author}{Hertel, R.}
\newblock \bibinfo{journal}{\bibinfo{title}{Ultrafast domain wall dynamics in
  magnetic nanotubes and nanowires}}.
\newblock {\emph{\JournalTitle{Journal of Physics: Condensed Matter}}}
  \textbf{\bibinfo{volume}{28}}, \bibinfo{pages}{483002}
  (\bibinfo{year}{2016}).
\newblock \urlprefix\url{https://doi.org/10.1088/0953-8984/28/48/483002}.

\bibitem{Sander_jpd2017}
\bibinfo{author}{Sander, D.} \emph{et~al.}
\newblock \bibinfo{journal}{\bibinfo{title}{The 2017 magnetism roadmap}}.
\newblock {\emph{\JournalTitle{Journal of Physics D: Applied Physics}}}
  \textbf{\bibinfo{volume}{50}}, \bibinfo{pages}{363001}
  (\bibinfo{year}{2017}).
\newblock \urlprefix\url{http://stacks.iop.org/0022-3727/50/i=36/a=363001}.

\bibitem{Sloika_jmmm2017}
\bibinfo{author}{Sloika, M.~I.}, \bibinfo{author}{Sheka, D.~D.},
  \bibinfo{author}{Kravchuk, V.~P.}, \bibinfo{author}{Pylypovskyi, O.~V.} \&
  \bibinfo{author}{Gaididei, Y.}
\newblock \bibinfo{journal}{\bibinfo{title}{Geometry induced phase transitions
  in magnetic spherical shell}}.
\newblock {\emph{\JournalTitle{J. Magn. Magn. Mater.}}}
  \textbf{\bibinfo{volume}{443}}, \bibinfo{pages}{404 -- 412}
  (\bibinfo{year}{2017}).
\newblock \doiprefix https://doi.org/10.1016/j.jmmm.2017.07.036.

\bibitem{Streubel_jpd2016}
\bibinfo{author}{Streubel, R.} \emph{et~al.}
\newblock \bibinfo{journal}{\bibinfo{title}{Magnetism in curved geometries}}.
\newblock {\emph{\JournalTitle{J. Phy. D: Appl. Phys.}}}
  \textbf{\bibinfo{volume}{49}}, \bibinfo{pages}{363001}
  (\bibinfo{year}{2016}).
\newblock \urlprefix\url{http://stacks.iop.org/0022-3727/49/i=36/a=363001}.

\bibitem{Hertel_spin2013}
\bibinfo{author}{Hertel, R.}
\newblock \bibinfo{journal}{\bibinfo{title}{Curvature-induced
  magnetochirality}}.
\newblock {\emph{\JournalTitle{Spin}}} \textbf{\bibinfo{volume}{03}},
  \bibinfo{pages}{1340009} (\bibinfo{year}{2013}).
\newblock \urlprefix\url{http://dx.doi.org/10.1142/S2010324713400092}.
\newblock \doiprefix 10.1142/S2010324713400092.

\bibitem{Sheka_prb2015}
\bibinfo{author}{Sheka, D.~D.}, \bibinfo{author}{Kravchuk, V.~P.},
  \bibinfo{author}{Yershov, K.~V.} \& \bibinfo{author}{Gaididei, Y.}
\newblock \bibinfo{journal}{\bibinfo{title}{Torsion-induced effects in magnetic
  nanowires}}.
\newblock {\emph{\JournalTitle{Phys. Rev. B}}} \textbf{\bibinfo{volume}{92}},
  \bibinfo{pages}{054417} (\bibinfo{year}{2015}).
\newblock \urlprefix\url{https://link.aps.org/doi/10.1103/PhysRevB.92.054417}.
\newblock \doiprefix 10.1103/PhysRevB.92.054417.

\bibitem{Fan_acsnano2009}
\bibinfo{author}{Fan, H.-M.} \emph{et~al.}
\newblock \bibinfo{journal}{\bibinfo{title}{Single-crystalline
  mfe$_2$o$_4$nanotubes/nanorings synthesized by thermal transformation process
  for biological applications}}.
\newblock {\emph{\JournalTitle{ACS Nano}}} \textbf{\bibinfo{volume}{3}},
  \bibinfo{pages}{2798--2808} (\bibinfo{year}{2009}).
\newblock \urlprefix\url{http://dx.doi.org/10.1021/nn9006797}.
\newblock \doiprefix 10.1021/nn9006797.

\bibitem{Xie_nanotec2008}
\bibinfo{author}{Xie, J.}, \bibinfo{author}{Chen, L.},
  \bibinfo{author}{Varadan, V.~K.}, \bibinfo{author}{Yancey, J.} \&
  \bibinfo{author}{Srivatsan, M.}
\newblock \bibinfo{journal}{\bibinfo{title}{The effects of functional magnetic
  nanotubes with incorporated nerve growth factor in neuronal differentiation
  of {PC12} cells}}.
\newblock {\emph{\JournalTitle{Nanotechnology}}} \textbf{\bibinfo{volume}{19}},
  \bibinfo{pages}{105101} (\bibinfo{year}{2008}).
\newblock \urlprefix\url{http://stacks.iop.org/0957-4484/19/i=10/a=105101}.
\newblock \doiprefix 10.1088/0957-4484/19/10/105101.

\bibitem{Pantarotto_chembio2013}
\bibinfo{author}{Pantarotto, D.} \emph{et~al.}
\newblock \bibinfo{journal}{\bibinfo{title}{Immunization with
  peptide-functionalized carbon nanotubes enhances virus-specific neutralizing
  antibody responses}}.
\newblock {\emph{\JournalTitle{Chemistry \& Biology}}}
  \textbf{\bibinfo{volume}{10}}, \bibinfo{pages}{961--966}
  (\bibinfo{year}{2013}).
\newblock \urlprefix\url{http://dx.doi.org/10.1016/j.chembiol.2003.09.011}.
\newblock \doiprefix 10.1016/j.chembiol.2003.09.011.

\bibitem{Krusin-Elbaum_nature2004}
\bibinfo{author}{Krusin-Elbaum, L.} \emph{et~al.}
\newblock \bibinfo{journal}{\bibinfo{title}{Room-temperature ferromagnetic
  nanotubes controlled by electron or hole doping}}.
\newblock {\emph{\JournalTitle{Nature}}} \textbf{\bibinfo{volume}{431}},
  \bibinfo{pages}{672--676} (\bibinfo{year}{2004}).
\newblock \urlprefix\url{http://dx.doi.org/10.1038/nature02970}.
\newblock \doiprefix 10.1038/nature02970.

\bibitem{Parkin_science2008}
\bibinfo{author}{Parkin, S. S.~P.}, \bibinfo{author}{Hayashi, M.} \&
  \bibinfo{author}{Thomas, L.}
\newblock \bibinfo{journal}{\bibinfo{title}{Magnetic domain-wall racetrack
  memory}}.
\newblock {\emph{\JournalTitle{Science}}} \textbf{\bibinfo{volume}{320}},
  \bibinfo{pages}{190--194} (\bibinfo{year}{2008}).
\newblock \urlprefix\url{http://science.sciencemag.org/content/320/5873/190}.
\newblock \doiprefix 10.1126/science.1145799.

\bibitem{Yan_apl2011}
\bibinfo{author}{Yan, M.}, \bibinfo{author}{Andreas, C.},
  \bibinfo{author}{Kákay, A.}, \bibinfo{author}{García-Sánchez, F.} \&
  \bibinfo{author}{Hertel, R.}
\newblock \bibinfo{journal}{\bibinfo{title}{Fast domain wall dynamics in
  magnetic nanotubes: Suppression of walker breakdown and cherenkov-like spin
  wave emission}}.
\newblock {\emph{\JournalTitle{Applied Physics Letters}}}
  \textbf{\bibinfo{volume}{99}}, \bibinfo{pages}{122505}
  (\bibinfo{year}{2011}).
\newblock \urlprefix\url{https://doi.org/10.1063/1.3643037}.
\newblock \doiprefix 10.1063/1.3643037.

\bibitem{Wang_prl2005}
\bibinfo{author}{Wang, Z.~K.} \emph{et~al.}
\newblock \bibinfo{journal}{\bibinfo{title}{Spin waves in nickel nanorings of
  large aspect ratio}}.
\newblock {\emph{\JournalTitle{Phys. Rev. Lett.}}}
  \textbf{\bibinfo{volume}{94}}, \bibinfo{pages}{137208}
  (\bibinfo{year}{2005}).
\newblock
  \urlprefix\url{https://link.aps.org/doi/10.1103/PhysRevLett.94.137208}.
\newblock \doiprefix 10.1103/PhysRevLett.94.137208.

\bibitem{Landeros_prb2009}
\bibinfo{author}{Landeros, P.}, \bibinfo{author}{Suarez, O.~J.},
  \bibinfo{author}{Cuchillo, A.} \& \bibinfo{author}{Vargas, P.}
\newblock \bibinfo{journal}{\bibinfo{title}{Equilibrium states and vortex
  domain wall nucleation in ferromagnetic nanotubes}}.
\newblock {\emph{\JournalTitle{Phys. Rev. B}}} \textbf{\bibinfo{volume}{79}},
  \bibinfo{pages}{024404} (\bibinfo{year}{2009}).
\newblock \urlprefix\url{http://link.aps.org/doi/10.1103/PhysRevB.79.024404}.
\newblock \doiprefix 10.1103/PhysRevB.79.024404.

\bibitem{Nielsch_jap2005}
\bibinfo{author}{Nielsch, K.}, \bibinfo{author}{Casta{\~n}o, F.~J.},
  \bibinfo{author}{Ross, C.~A.} \& \bibinfo{author}{Krishnan, R.}
\newblock \bibinfo{journal}{\bibinfo{title}{Magnetic properties of
  template-synthesized cobalt polymer composite nanotubes}}.
\newblock {\emph{\JournalTitle{J. Appl. Phys.}}} \textbf{\bibinfo{volume}{98}}
  (\bibinfo{year}{2005}).
\newblock
  \urlprefix\url{http://scitation.aip.org/content/aip/journal/jap/98/3/10.1063/1.2005384}.
\newblock \doiprefix 10.1063/1.2005384.

\bibitem{Kohli_jppc2010}
\bibinfo{author}{Kohli, S.} \emph{et~al.}
\newblock \bibinfo{journal}{\bibinfo{title}{Template-assisted chemical vapor
  deposited spinel ferrite nanotubes}}.
\newblock {\emph{\JournalTitle{J. Phys. Chem. C}}}
  \textbf{\bibinfo{volume}{114}}, \bibinfo{pages}{19557--19561}
  (\bibinfo{year}{2010}).
\newblock \urlprefix\url{https://doi.org/10.1021/jp101099u}.
\newblock \doiprefix 10.1021/jp101099u.

\bibitem{Siu_apl2004}
\bibinfo{author}{Sui, Y.~C.}, \bibinfo{author}{Skomski, R.},
  \bibinfo{author}{Sorge, K.~D.} \& \bibinfo{author}{Sellmyer, D.~J.}
\newblock \bibinfo{journal}{\bibinfo{title}{Nanotube magnetism}}.
\newblock {\emph{\JournalTitle{Appl. Phys. Lett.}}}
  \textbf{\bibinfo{volume}{84}}, \bibinfo{pages}{1525--1527}
  (\bibinfo{year}{2004}).
\newblock
  \urlprefix\url{http://scitation.aip.org/content/aip/journal/apl/84/9/10.1063/1.1655692}.
\newblock \doiprefix 10.1063/1.1655692.

\bibitem{Xu_jpd2008}
\bibinfo{author}{Xu, Y.}, \bibinfo{author}{Xue, D.~S.}, \bibinfo{author}{Fu,
  J.~L.}, \bibinfo{author}{Gao, D.~Q.} \& \bibinfo{author}{Gao, B.}
\newblock \bibinfo{journal}{\bibinfo{title}{Synthesis, characterization and
  magnetic properties of {Fe} nanotubes}}.
\newblock {\emph{\JournalTitle{J. Phys. D: Appl. Phys.}}}
  \textbf{\bibinfo{volume}{41}}, \bibinfo{pages}{215010}
  (\bibinfo{year}{2008}).
\newblock \urlprefix\url{http://stacks.iop.org/0022-3727/41/i=21/a=215010}.
\newblock \doiprefix 10.1088/0022-3727/41/21/215010.

\bibitem{Proenca_book2015}
\bibinfo{author}{Proenca, M.}, \bibinfo{author}{Sousa, C.},
  \bibinfo{author}{Ventura, J.} \& \bibinfo{author}{Araújo, J.}
\newblock \bibinfo{title}{Electrochemical synthesis and magnetism of magnetic
  nanotubes}.
\newblock In \bibinfo{editor}{Vázquez, M.} (ed.)
  \emph{\bibinfo{booktitle}{Magnetic Nano- and Microwires}}, Woodhead
  Publishing Series in Electronic and Optical Materials, \bibinfo{pages}{727 --
  781} (\bibinfo{publisher}{Woodhead Publishing}, \bibinfo{year}{2015}).
\newblock
  \urlprefix\url{https://www.sciencedirect.com/science/article/pii/B9780081001646000242}.
\newblock \doiprefix https://doi.org/10.1016/B978-0-08-100164-6.00024-2.

\bibitem{Wang_cc2010}
\bibinfo{author}{Wang, Q.}, \bibinfo{author}{Geng, B.}, \bibinfo{author}{Wang,
  S.}, \bibinfo{author}{Ye, Y.} \& \bibinfo{author}{Tao, B.}
\newblock \bibinfo{journal}{\bibinfo{title}{Modified kirkendall effect for
  fabrication of magnetic nanotubes}}.
\newblock {\emph{\JournalTitle{Chem. Commun.}}} \textbf{\bibinfo{volume}{46}},
  \bibinfo{pages}{1899--1901} (\bibinfo{year}{2010}).
\newblock \urlprefix\url{http://dx.doi.org/10.1039/B922134D}.
\newblock \doiprefix 10.1039/B922134D.

\bibitem{Harada_pccp2010}
\bibinfo{author}{Harada, T.}, \bibinfo{author}{Simeon, F.},
  \bibinfo{author}{Vander~Sande, J.~B.} \& \bibinfo{author}{Hatton, T.~A.}
\newblock \bibinfo{journal}{\bibinfo{title}{Formation of magnetic nanotubes by
  the cooperative self-assembly of chiral amphiphilic molecules and fe$_3$o$_4$
  nanoparticles}}.
\newblock {\emph{\JournalTitle{Phys. Chem. Chem. Phys.}}}
  \textbf{\bibinfo{volume}{12}}, \bibinfo{pages}{11938--11943}
  (\bibinfo{year}{2010}).
\newblock \urlprefix\url{http://dx.doi.org/10.1039/C0CP00533A}.
\newblock \doiprefix 10.1039/C0CP00533A.

\bibitem{Ye_crssms2012}
\bibinfo{author}{Ye, Y.} \& \bibinfo{author}{Geng, B.}
\newblock \bibinfo{journal}{\bibinfo{title}{Magnetic nanotubes: Synthesis,
  properties, and applications}}.
\newblock {\emph{\JournalTitle{Critical Reviews in Solid State and Materials
  Sciences}}} \textbf{\bibinfo{volume}{37}}, \bibinfo{pages}{75--93}
  (\bibinfo{year}{2012}).
\newblock \urlprefix\url{https://doi.org/10.1080/10408436.2011.613491}.
\newblock \doiprefix 10.1080/10408436.2011.613491.

\bibitem{Streubel_NanoLetters2014}
\bibinfo{author}{Streubel, R.} \emph{et~al.}
\newblock \bibinfo{journal}{\bibinfo{title}{Imaging of buried 3d magnetic
  rolled-up nanomembranes}}.
\newblock {\emph{\JournalTitle{Nano Letters}}} \textbf{\bibinfo{volume}{14}},
  \bibinfo{pages}{3981--3986} (\bibinfo{year}{2014}).
\newblock \urlprefix\url{http://dx.doi.org/10.1021/nl501333h}.
\newblock \doiprefix 10.1021/nl501333h.

\bibitem{Liu_jacs2005}
\bibinfo{author}{Liu, L.} \emph{et~al.}
\newblock \bibinfo{journal}{\bibinfo{title}{Single crystalline magnetite
  nanotubes}}.
\newblock {\emph{\JournalTitle{J. Am. Chem. Soc}}}
  \textbf{\bibinfo{volume}{127}}, \bibinfo{pages}{6} (\bibinfo{year}{2005}).
\newblock \urlprefix\url{http://dx.doi.org/10.1021/ja0445239}.
\newblock \doiprefix 10.1021/ja0445239.

\bibitem{Sotnikov_prb2017}
\bibinfo{author}{Sotnikov, O.~M.}, \bibinfo{author}{Mazurenko, V.~V.} \&
  \bibinfo{author}{Katanin, A.~A.}
\newblock \bibinfo{journal}{\bibinfo{title}{Monte carlo study of magnetic
  nanoparticles adsorbed on halloysite
  ${\mathrm{al}}_{2}{\mathrm{si}}_{2}{\mathrm{o}}_{5}{(\mathrm{OH})}_{4}$
  nanotubes}}.
\newblock {\emph{\JournalTitle{Phys. Rev. B}}} \textbf{\bibinfo{volume}{96}},
  \bibinfo{pages}{224404} (\bibinfo{year}{2017}).
\newblock \urlprefix\url{https://link.aps.org/doi/10.1103/PhysRevB.96.224404}.
\newblock \doiprefix 10.1103/PhysRevB.96.224404.

\bibitem{Jang_advmat2003}
\bibinfo{author}{Jang, J.} \& \bibinfo{author}{Yoon, H.}
\newblock \bibinfo{journal}{\bibinfo{title}{Fabrication of magnetic carbon
  nanotubes using a metal-impregnated polymer precursor}}.
\newblock {\emph{\JournalTitle{Adv. Mater.}}} \textbf{\bibinfo{volume}{15}},
  \bibinfo{pages}{2088--2091} (\bibinfo{year}{2003}).
\newblock \urlprefix\url{http://dx.doi.org/10.1002/adma.200305296}.
\newblock \doiprefix 10.1002/adma.200305296.

\bibitem{Gao_jpcb2006}
\bibinfo{author}{Gao, C.}, \bibinfo{author}{Li, W.}, \bibinfo{author}{Morimoto,
  H.}, \bibinfo{author}{Nagaoka, Y.} \& \bibinfo{author}{Maekawa, T.}
\newblock \bibinfo{journal}{\bibinfo{title}{Magnetic carbon nanotubes:
  Synthesis by electrostatic self-assembly approach and application in
  biomanipulations}}.
\newblock {\emph{\JournalTitle{Journal of Physical Chemistry B}}}
  \textbf{\bibinfo{volume}{110}}, \bibinfo{pages}{7213--7220}
  (\bibinfo{year}{2006}).
\newblock \urlprefix\url{http://dx.doi.org/10.1021/jp0602474}.
\newblock \doiprefix 10.1021/jp0602474.

\bibitem{Lamanna_nanoscale2013}
\bibinfo{author}{Lamanna, G.} \emph{et~al.}
\newblock \bibinfo{journal}{\bibinfo{title}{Endowing carbon nanotubes with
  superparamagnetic properties: applications for cell labeling{,} mri cell
  tracking and magnetic manipulations}}.
\newblock {\emph{\JournalTitle{Nanoscale}}} \textbf{\bibinfo{volume}{5}},
  \bibinfo{pages}{4412--4421} (\bibinfo{year}{2013}).
\newblock \urlprefix\url{http://dx.doi.org/10.1039/C3NR00636K}.
\newblock \doiprefix 10.1039/C3NR00636K.

\bibitem{Kim_jpcc2010}
\bibinfo{author}{Kim, I.~T.} \emph{et~al.}
\newblock \bibinfo{journal}{\bibinfo{title}{Synthesis, characterization, and
  alignment of magnetic carbon nanotubes tethered with maghemite
  nanoparticles}}.
\newblock {\emph{\JournalTitle{J. Phys. Chem. C}}}
  \textbf{\bibinfo{volume}{114}}, \bibinfo{pages}{6944--6951}
  (\bibinfo{year}{2010}).
\newblock \urlprefix\url{http://dx.doi.org/10.1021/jp9118925}.
\newblock \doiprefix 10.1021/jp9118925.

\bibitem{Bogani_Angewandte2009}
\bibinfo{author}{Bogani, L.} \emph{et~al.}
\newblock \bibinfo{journal}{\bibinfo{title}{Single-molecule-magnet
  carbon-nanotube hybrids}}.
\newblock {\emph{\JournalTitle{Angewandte Chemie - International Edition}}}
  \textbf{\bibinfo{volume}{48}}, \bibinfo{pages}{746--750}
  (\bibinfo{year}{2009}).
\newblock \urlprefix\url{http://dx.doi.org/10.1002/anie.200804967}.
\newblock \doiprefix 10.1002/anie.200804967.

\bibitem{Cowburn_prl1999}
\bibinfo{author}{Cowburn, R.~P.}, \bibinfo{author}{Koltsov, D.~K.},
  \bibinfo{author}{Adeyeye, A.~O.}, \bibinfo{author}{Welland, M.~E.} \&
  \bibinfo{author}{Tricker, D.~M.}
\newblock \bibinfo{journal}{\bibinfo{title}{Single-domain circular
  nanomagnets}}.
\newblock {\emph{\JournalTitle{Phys. Rev. Lett.}}}
  \textbf{\bibinfo{volume}{83}}, \bibinfo{pages}{1042--1045}
  (\bibinfo{year}{1999}).
\newblock \urlprefix\url{https://link.aps.org/doi/10.1103/PhysRevLett.83.1042}.
\newblock \doiprefix 10.1103/PhysRevLett.83.1042.

\bibitem{Chen_jmmm2007}
\bibinfo{author}{Chen, A.}, \bibinfo{author}{Usov, N.},
  \bibinfo{author}{Blanco, J.} \& \bibinfo{author}{Gonzalez, J.}
\newblock \bibinfo{journal}{\bibinfo{title}{Equilibrium magnetization states in
  magnetic nanotubes and their evolution in external magnetic field}}.
\newblock {\emph{\JournalTitle{J. Magn. Magn. Mater.}}}
  \textbf{\bibinfo{volume}{316}}, \bibinfo{pages}{e317 -- e319}
  (\bibinfo{year}{2007}).
\newblock \doiprefix 10.1016/j.jmmm.2007.02.132.

\bibitem{Escrig_jmmm2007}
\bibinfo{author}{Escrig, J.}, \bibinfo{author}{Landeros, P.},
  \bibinfo{author}{Altbir, D.}, \bibinfo{author}{Vogel, E.} \&
  \bibinfo{author}{Vargas, P.}
\newblock \bibinfo{journal}{\bibinfo{title}{Phase diagrams of magnetic
  nanotubes}}.
\newblock {\emph{\JournalTitle{J. Magn. Magn. Mater.}}}
  \textbf{\bibinfo{volume}{308}}, \bibinfo{pages}{233 -- 237}
  (\bibinfo{year}{2007}).
\newblock \doiprefix 10.1016/j.jmmm.2006.05.019.

\bibitem{Raabe_jap2000}
\bibinfo{author}{Raabe, J.} \emph{et~al.}
\newblock \bibinfo{journal}{\bibinfo{title}{Magnetization pattern of
  ferromagnetic nanodisks}}.
\newblock {\emph{\JournalTitle{J. App. Phys.}}} \textbf{\bibinfo{volume}{88}},
  \bibinfo{pages}{4437--4439} (\bibinfo{year}{2000}).
\newblock \urlprefix\url{http://aip.scitation.org/doi/abs/10.1063/1.1289216}.
\newblock \doiprefix 10.1063/1.1289216.

\bibitem{Biziere_Nanolett2013}
\bibinfo{author}{Biziere, N.} \emph{et~al.}
\newblock \bibinfo{journal}{\bibinfo{title}{Imaging the fine structure of a
  magnetic domain wall in a ni nanocylinder}}.
\newblock {\emph{\JournalTitle{Nano Letters}}} \textbf{\bibinfo{volume}{13}},
  \bibinfo{pages}{2053--2057} (\bibinfo{year}{2013}).
\newblock \urlprefix\url{http://dx.doi.org/10.1021/nl400317j}.
\newblock \doiprefix 10.1021/nl400317j.

\bibitem{Weber_Nanolett2012}
\bibinfo{author}{Weber, D.~P.} \emph{et~al.}
\newblock \bibinfo{journal}{\bibinfo{title}{Cantilever magnetometry of
  individual ni nanotubes}}.
\newblock {\emph{\JournalTitle{Nano Letters}}} \textbf{\bibinfo{volume}{12}},
  \bibinfo{pages}{6139--6144} (\bibinfo{year}{2012}).
\newblock \doiprefix 10.1021/nl302950u.

\bibitem{Wyss_prb2017}
\bibinfo{author}{Wyss, M.} \emph{et~al.}
\newblock \bibinfo{journal}{\bibinfo{title}{Imaging magnetic vortex
  configurations in ferromagnetic nanotubes}}.
\newblock {\emph{\JournalTitle{Phys. Rev. B}}} \textbf{\bibinfo{volume}{96}},
  \bibinfo{pages}{024423} (\bibinfo{year}{2017}).
\newblock \urlprefix\url{https://link.aps.org/doi/10.1103/PhysRevB.96.024423}.
\newblock \doiprefix 10.1103/PhysRevB.96.024423.

\bibitem{Vasyukov_NanoLett2018}
\bibinfo{author}{Vasyukov, D.} \emph{et~al.}
\newblock \bibinfo{journal}{\bibinfo{title}{Imaging stray magnetic field of
  individual ferromagnetic nanotubes}}.
\newblock {\emph{\JournalTitle{Nano Letters}}} \textbf{\bibinfo{volume}{18}},
  \bibinfo{pages}{964} (\bibinfo{year}{2018}).
\newblock \urlprefix\url{https://dx.doi.org/10.1021/acs.nanolett.7b04386}.
\newblock \doiprefix 10.1021/acs.nanolett.7b04386.

\bibitem{Mehlin_prb2018}
\bibinfo{author}{Mehlin, A.} \emph{et~al.}
\newblock \bibinfo{journal}{\bibinfo{title}{{Observation of end-vortex
  nucleation in individual ferromagnetic nanotubes}}}.
\newblock {\emph{\JournalTitle{Phys. Rev. B}}} \textbf{\bibinfo{volume}{97}},
  \bibinfo{pages}{134422} (\bibinfo{year}{2018}).
\newblock \urlprefix\url{https://link.aps.org/doi/10.1103/PhysRevB.97.134422}.
\newblock \doiprefix 10.1103/PhysRevB.97.134422.

\bibitem{Buchter_prl2013}
\bibinfo{author}{Buchter, A.} \emph{et~al.}
\newblock \bibinfo{journal}{\bibinfo{title}{Reversal mechanism of an individual
  ni nanotube simultaneously studied by torque and squid magnetometry}}.
\newblock {\emph{\JournalTitle{Phys. Rev. Lett.}}}
  \textbf{\bibinfo{volume}{111}}, \bibinfo{pages}{067202}
  (\bibinfo{year}{2013}).
\newblock \doiprefix 10.1103/PhysRevLett.111.067202.

\bibitem{Yamasaki_prl2003}
\bibinfo{author}{Yamasaki, A.}, \bibinfo{author}{Wulfhekel, W.},
  \bibinfo{author}{Hertel, R.}, \bibinfo{author}{Suga, S.} \&
  \bibinfo{author}{Kirschner, J.}
\newblock \bibinfo{journal}{\bibinfo{title}{Direct observation of the
  single-domain limit of fe nanomagnets by spin-polarized scanning tunneling
  spectroscopy}}.
\newblock {\emph{\JournalTitle{Phys. Rev. Lett.}}}
  \textbf{\bibinfo{volume}{91}}, \bibinfo{pages}{127201}
  (\bibinfo{year}{2003}).
\newblock
  \urlprefix\url{http://link.aps.org/doi/10.1103/PhysRevLett.91.127201}.
\newblock \doiprefix 10.1103/PhysRevLett.91.127201.

\bibitem{Zimmermann_nanolett2018}
\bibinfo{author}{Zimmermann, M.} \emph{et~al.}
\newblock \bibinfo{journal}{\bibinfo{title}{Origin and manipulation of stable
  vortex ground states in permalloy nanotubes}}.
\newblock {\emph{\JournalTitle{Nano Letters}}} \textbf{\bibinfo{volume}{18}},
  \bibinfo{pages}{2828--2834} (\bibinfo{year}{2018}).
\newblock \urlprefix\url{https://doi.org/10.1021/acs.nanolett.7b05222}.
\newblock \doiprefix 10.1021/acs.nanolett.7b05222.

\bibitem{Lebib_jap2001}
\bibinfo{author}{Lebib, A.}, \bibinfo{author}{Li, S.~P.},
  \bibinfo{author}{Natali, M.} \& \bibinfo{author}{Chen, Y.}
\newblock \bibinfo{journal}{\bibinfo{title}{Size and thickness dependencies of
  magnetization reversal in co dot arrays}}.
\newblock {\emph{\JournalTitle{J. Appl. Phys.}}} \textbf{\bibinfo{volume}{89}},
  \bibinfo{pages}{3892--3896} (\bibinfo{year}{2001}).
\newblock
  \urlprefix\url{http://scitation.aip.org/content/aip/journal/jap/89/7/10.1063/1.1355282}.
\newblock \doiprefix 10.1063/1.1355282.

\bibitem{Allende_ejpb2008}
\bibinfo{author}{Allende, S.}, \bibinfo{author}{Escrig, J.},
  \bibinfo{author}{Altbir, D.}, \bibinfo{author}{Salcedo, E.} \&
  \bibinfo{author}{Bahiana, M.}
\newblock \bibinfo{journal}{\bibinfo{title}{Angular dependence of the
  transverse and vortex modes in magnetic nanotubes}}.
\newblock {\emph{\JournalTitle{Eur. Phys. J. B}}}
  \textbf{\bibinfo{volume}{66}}, \bibinfo{pages}{37--40}
  (\bibinfo{year}{2008}).
\newblock \urlprefix\url{http://dx.doi.org/10.1140/epjb/e2008-00385-4}.
\newblock \doiprefix 10.1140/epjb/e2008-00385-4.

\bibitem{Albuquerque_prl2002}
\bibinfo{author}{d'Albuquerque~e Castro, J.}, \bibinfo{author}{Altbir, D.},
  \bibinfo{author}{Retamal, J.~C.} \& \bibinfo{author}{Vargas, P.}
\newblock \bibinfo{journal}{\bibinfo{title}{Scaling approach to the magnetic
  phase diagram of nanosized systems}}.
\newblock {\emph{\JournalTitle{Phys. Rev. Lett.}}}
  \textbf{\bibinfo{volume}{88}}, \bibinfo{pages}{237202}
  (\bibinfo{year}{2002}).
\newblock
  \urlprefix\url{http://link.aps.org/doi/10.1103/PhysRevLett.88.237202}.
\newblock \doiprefix 10.1103/PhysRevLett.88.237202.

\bibitem{Landeros_prb2005}
\bibinfo{author}{Landeros, P.} \emph{et~al.}
\newblock \bibinfo{journal}{\bibinfo{title}{Scaling relations for magnetic
  nanoparticles}}.
\newblock {\emph{\JournalTitle{Phys. Rev. B}}} \textbf{\bibinfo{volume}{71}},
  \bibinfo{pages}{094435} (\bibinfo{year}{2005}).
\newblock \urlprefix\url{http://link.aps.org/doi/10.1103/PhysRevB.71.094435}.
\newblock \doiprefix 10.1103/PhysRevB.71.094435.

\bibitem{Mi_jmmm2010}
\bibinfo{author}{Mi, B.-Z.}, \bibinfo{author}{Wang, H.-Y.} \&
  \bibinfo{author}{Zhou, Y.-S.}
\newblock \bibinfo{journal}{\bibinfo{title}{Theoretical investigations of
  magnetic properties of ferromagnetic single-walled nanotubes}}.
\newblock {\emph{\JournalTitle{J. Magn. Magn. Mater.}}}
  \textbf{\bibinfo{volume}{322}}, \bibinfo{pages}{952 -- 958}
  (\bibinfo{year}{2010}).
\newblock
  \urlprefix\url{http://www.sciencedirect.com/science/article/pii/S0304885309011366}.
\newblock \doiprefix 10.1016/j.jmmm.2009.11.030.

\bibitem{Mi_jmmm2016}
\bibinfo{author}{Mi, B.-Z.}, \bibinfo{author}{Zhai, L.-J.} \&
  \bibinfo{author}{Hua, L.-L.}
\newblock \bibinfo{journal}{\bibinfo{title}{Magnon specific heat and free
  energy of heisenberg ferromagnetic single-walled nanotubes: Green's function
  approach}}.
\newblock {\emph{\JournalTitle{J. Magn. Magn. Mater.}}}
  \textbf{\bibinfo{volume}{398}}, \bibinfo{pages}{160 -- 166}
  (\bibinfo{year}{2016}).
\newblock
  \urlprefix\url{http://www.sciencedirect.com/science/article/pii/S0304885315305692}.
\newblock \doiprefix https://doi.org/10.1016/j.jmmm.2015.09.016.

\bibitem{Mi_phyb2016}
\bibinfo{author}{Mi, B.-Z.}
\newblock \bibinfo{journal}{\bibinfo{title}{Spin wave dynamics in heisenberg
  ferromagnetic/antiferromagnetic single-walled nanotubes}}.
\newblock {\emph{\JournalTitle{Physica B: Condensed Matter}}}
  \textbf{\bibinfo{volume}{497}}, \bibinfo{pages}{23 -- 30}
  (\bibinfo{year}{2016}).
\newblock
  \urlprefix\url{http://www.sciencedirect.com/science/article/pii/S0921452616302216}.
\newblock \doiprefix https://doi.org/10.1016/j.physb.2016.05.029.

\bibitem{Lehtinen_prl2004}
\bibinfo{author}{Lehtinen, P.~O.}, \bibinfo{author}{Foster, A.~S.},
  \bibinfo{author}{Ma, Y.}, \bibinfo{author}{Krasheninnikov, A.~V.} \&
  \bibinfo{author}{Nieminen, R.~M.}
\newblock \bibinfo{journal}{\bibinfo{title}{Irradiation-induced magnetism in
  graphite: A density functional study}}.
\newblock {\emph{\JournalTitle{Phys. Rev. Lett.}}}
  \textbf{\bibinfo{volume}{93}}, \bibinfo{pages}{187202}
  (\bibinfo{year}{2004}).
\newblock \doiprefix 10.1103/PhysRevLett.93.187202.

\bibitem{Shimada_prb2011}
\bibinfo{author}{Shimada, T.}, \bibinfo{author}{Ishii, Y.} \&
  \bibinfo{author}{Kitamura, T.}
\newblock \bibinfo{journal}{\bibinfo{title}{Ab initio study of ferromagnetic
  single-wall nickel nanotubes}}.
\newblock {\emph{\JournalTitle{Phys. Rev. B}}} \textbf{\bibinfo{volume}{84}},
  \bibinfo{pages}{165452} (\bibinfo{year}{2011}).
\newblock \urlprefix\url{https://link.aps.org/doi/10.1103/PhysRevB.84.165452}.
\newblock \doiprefix 10.1103/PhysRevB.84.165452.

\bibitem{Shimada_nanolett2013}
\bibinfo{author}{Shimada, T.}, \bibinfo{author}{Okuno, J.} \&
  \bibinfo{author}{Kitamura, T.}
\newblock \bibinfo{journal}{\bibinfo{title}{Chiral selectivity of unusual
  helimagnetic transition in iron nanotubes: Chirality makes quantum
  helimagnets}}.
\newblock {\emph{\JournalTitle{Nano Letters}}} \textbf{\bibinfo{volume}{13}},
  \bibinfo{pages}{2792--2797} (\bibinfo{year}{2013}).
\newblock \urlprefix\url{https://doi.org/10.1021/nl401047z}.
\newblock \doiprefix 10.1021/nl401047z.

\bibitem{Konstantinova_jmmm2008}
\bibinfo{author}{Konstantinova, E.}
\newblock \bibinfo{journal}{\bibinfo{title}{Theoretical simulations of magnetic
  nanotubes using monte carlo method}}.
\newblock {\emph{\JournalTitle{J. Magn. Magn. Mater.}}}
  \textbf{\bibinfo{volume}{320}}, \bibinfo{pages}{2721 -- 2729}
  (\bibinfo{year}{2008}).
\newblock \doiprefix 10.1016/j.jmmm.2008.06.007.

\bibitem{Chien_phystoday2007}
\bibinfo{author}{Chien, C.~L.}, \bibinfo{author}{Zhu, F.~Q.} \&
  \bibinfo{author}{Zhu, J.~G.}
\newblock \bibinfo{journal}{\bibinfo{title}{Patterned nanomagnets}}.
\newblock {\emph{\JournalTitle{Physics Today}}} \textbf{\bibinfo{volume}{60}},
  \bibinfo{pages}{40} (\bibinfo{year}{2007}).
\newblock \urlprefix\url{http://dx.doi.org/10.1063/1.2754602}.
\newblock \doiprefix 10.1063/1.2754602.

\bibitem{Salinas_jsnm2012}
\bibinfo{author}{Salinas, H.~D.} \& \bibinfo{author}{Restrepo, J.}
\newblock \bibinfo{journal}{\bibinfo{title}{Influence of the competition
  between dipolar and exchange interactions on the magnetic structure of
  single-wall nanocylinders. monte carlo simulation}}.
\newblock {\emph{\JournalTitle{J. Supercond. Nov. Magn.}}}
  \textbf{\bibinfo{volume}{25}}, \bibinfo{pages}{2217--2221}
  (\bibinfo{year}{2012}).
\newblock \urlprefix\url{http://dx.doi.org/10.1007/s10948-012-1651-9}.
\newblock \doiprefix 10.1007/s10948-012-1651-9.

\bibitem{Masotti_ijms2013}
\bibinfo{author}{Masotti, A.} \& \bibinfo{author}{Caporali, A.}
\newblock \bibinfo{journal}{\bibinfo{title}{Preparation of magnetic carbon
  nanotubes for biomedical and biotechnological applications}}.
\newblock {\emph{\JournalTitle{Int. J. Mol. Sci.}}}
  \textbf{\bibinfo{volume}{14}}, \bibinfo{pages}{24619} (\bibinfo{year}{2013}).
\newblock \urlprefix\url{http://www.mdpi.com/1422-0067/14/12/24619}.
\newblock \doiprefix 10.3390/ijms141224619.

\bibitem{Li-Ying}
\bibinfo{author}{Li, Y.}, \bibinfo{author}{Wang, T.} \& \bibinfo{author}{Li,
  Y.}
\newblock \bibinfo{journal}{\bibinfo{title}{The influence of dipolar
  interaction on magnetic properties in nanomagnets with different shapes}}.
\newblock {\emph{\JournalTitle{Phys. Status Solidi B}}}
  \textbf{\bibinfo{volume}{247}}, \bibinfo{pages}{1237--1241}
  (\bibinfo{year}{2010}).
\newblock \urlprefix\url{http://dx.doi.org/10.1002/pssb.200945471}.
\newblock \doiprefix 10.1002/pssb.200945471.

\bibitem{Fisher}
\bibinfo{author}{Fisher, M.~E.} \& \bibinfo{author}{Barber, M.~N.}
\newblock \bibinfo{journal}{\bibinfo{title}{Scaling theory for finite-size
  effects in the critical region}}.
\newblock {\emph{\JournalTitle{Phys. Rev. Lett.}}}
  \textbf{\bibinfo{volume}{28}}, \bibinfo{pages}{1516--1519}
  (\bibinfo{year}{1972}).
\newblock \urlprefix\url{http://link.aps.org/doi/10.1103/PhysRevLett.28.1516}.
\newblock \doiprefix 10.1103/PhysRevLett.28.1516.

\bibitem{Velasquez_prb2015}
\bibinfo{author}{Vel\'asquez, E.~A.}, \bibinfo{author}{Mazo-Zuluaga, J.},
  \bibinfo{author}{Vargas, P.} \& \bibinfo{author}{Mej\'{\i}a-L\'opez, J.}
\newblock \bibinfo{journal}{\bibinfo{title}{Bridging the gap between discrete
  and continuous magnetic models in the scaling approach}}.
\newblock {\emph{\JournalTitle{Phys. Rev. B}}} \textbf{\bibinfo{volume}{91}},
  \bibinfo{pages}{134418} (\bibinfo{year}{2015}).
\newblock \urlprefix\url{https://link.aps.org/doi/10.1103/PhysRevB.91.134418}.
\newblock \doiprefix 10.1103/PhysRevB.91.134418.

\bibitem{Escrig_jmmm2007b}
\bibinfo{author}{Escrig, J.}, \bibinfo{author}{Landeros, P.},
  \bibinfo{author}{Altbir, D.} \& \bibinfo{author}{Vogel, E.}
\newblock \bibinfo{journal}{\bibinfo{title}{Effect of anisotropy in magnetic
  nanotubes}}.
\newblock {\emph{\JournalTitle{J. Magn. Magn. Mater.}}}
  \textbf{\bibinfo{volume}{310}}, \bibinfo{pages}{2448 -- 2450}
  (\bibinfo{year}{2007}).
\newblock \doiprefix 10.1016/j.jmmm.2006.10.910.

\bibitem{Chen_jap2010}
\bibinfo{author}{Chen, A.~P.}, \bibinfo{author}{Guslienko, K.~Y.} \&
  \bibinfo{author}{Gonzalez, J.}
\newblock \bibinfo{journal}{\bibinfo{title}{Magnetization configurations and
  reversal of thin magnetic nanotubes with uniaxial anisotropy}}.
\newblock {\emph{\JournalTitle{Journal of Applied Physics}}}
  \textbf{\bibinfo{volume}{108}}, \bibinfo{pages}{083920}
  (\bibinfo{year}{2010}).
\newblock \doiprefix 10.1063/1.3488630.

\bibitem{Chen_jap2011}
\bibinfo{author}{Chen, A.-P.}, \bibinfo{author}{Gonzalez, J.~M.} \&
  \bibinfo{author}{Guslienko, K.~Y.}
\newblock \bibinfo{journal}{\bibinfo{title}{Magnetization configurations and
  reversal of magnetic nanotubes with opposite chiralities of the end
  domains}}.
\newblock {\emph{\JournalTitle{Journal of Applied Physics}}}
  \textbf{\bibinfo{volume}{109}}, \bibinfo{pages}{073923}
  (\bibinfo{year}{2011}).
\newblock \doiprefix 10.1063/1.3562190.

\bibitem{Lopez_jmmm2012}
\bibinfo{author}{López-López, J.}, \bibinfo{author}{Cortés-Ortuño, D.} \&
  \bibinfo{author}{Landeros, P.}
\newblock \bibinfo{journal}{\bibinfo{title}{Role of anisotropy on the domain
  wall properties of ferromagnetic nanotubes}}.
\newblock {\emph{\JournalTitle{J. Magn. Magn. Mater.}}}
  \textbf{\bibinfo{volume}{324}}, \bibinfo{pages}{2024 -- 2029}
  (\bibinfo{year}{2012}).
\newblock \doiprefix https://doi.org/10.1016/j.jmmm.2012.01.040.

\end{thebibliography}

\section*{Acknowledgements}
O. I. acknowledges financial support form the Spanish MINECO (MAT2015-68772), Catalan DURSI (2014SGR220) and European Union FEDER Funds (Una manera de hacer Europa), also CSUC for supercomputer facilities.
H.D.S and J.R acknowledge financial support from the Colciencias "Beca de Doctorados nacionales, convocatoria 727",  the project CODI-UdeA 2016-10085. J.R acknowledges UdeA for the exclusive dedication program.

\section*{Author contributions}
H.D.S. performed the MC simulations under supervision of J.R. and O.I. O.I. conceived and performed the comparison to analytical calculations. Discussion and interpretation of the results was performed by all authors. All authors participated in the writing and revising of the manuscript. 

\section*{Additional information}
\textbf{Competing interests} The authors declare no competing interests. 

\end{document}